\documentclass[a4paper,twoside,10pt]{letter}
\usepackage{graphicx,saj,multicol,subeqnarray}

\def\papertype{Original Scientific Paper}

\begin{document}
\baselineskip=3.1truemm
\columnsep=.5truecm
\newenvironment{lefteqnarray}{\arraycolsep=0pt\begin{eqnarray}}
{\end{eqnarray}\protect\aftergroup\ignorespaces}
\newenvironment{lefteqnarray*}{\arraycolsep=0pt\begin{eqnarray*}}
{\end{eqnarray*}\protect\aftergroup\ignorespaces}
\newenvironment{leftsubeqnarray}{\arraycolsep=0pt\begin{subeqnarray}}
{\end{subeqnarray}\protect\aftergroup\ignorespaces}
%

% Running titles

\markboth{\eightrm Photometric and spectroscopic study of 5 pre-main sequence stars in the vicinity of NGC 7129} {\eightrm E. SEMKOV, S. IBRYAMOV AND S. PENEVA}

{\ }

\publ

\type

{\ }

% Title

\title{Photometric and spectroscopic study of 5 pre-main sequence stars in the vicinity of NGC 7129}

% Authors

\authors{E. H. Semkov$^{1}$, S. I. Ibryamov$^{2}$ and S. P. Peneva$^{1}$}

\vskip3mm

% Address

\address{$^1$Institute of Astronomy and National Astronomical Observatory, Bulgarian Academy of Sciences, 72, Tsarigradsko Shose Blvd., 1784 Sofia, Bulgaria}

\Email{esemkov}{astro.bas.bg, speneva@astro.bas.bg}

\address{$^2$Department of Physics and Astronomy, Faculty of Natural Sciences, University of Shumen, 115, Universitetska Str., 9712 Shumen, Bulgaria}

\Email{sibryamov}{shu.bg}

% Received and Accepted dates

\dates{.}{.}

% Abstract

\summary{We present results from long-term optical photometric and spectroscopic observations of 5 pre-main sequence stars, located in the vicinity of the bight nebula NGC 7129. 
We obtained $UBVRI$ photometric observations in the field centered on the star V391 Cep, north-west of the bright nebula NGC 7129. 
Our multicolor CCD observations spanned the period from February 1998 to November 2016. 
At the time of our photometric monitoring, a total of thirteen medium-resolution optical spectra of the stars were obtained. 
The results from our photometric study show that all stars exhibit strong variability in all optical passbands. 
Long-term light curves of the five stars indicate the typical of classical T Tauri stars variations in brightness with large amplitudes.
We did not find any reliable periodicity in the brightness variations of all five stars.
The results from spectral observations showed that all studied stars can be classified as a classical T Tauri stars with reach emission line spectra and strong variability in the profiles and intensity of emission lines.}

% Keywords (see keywords.pdf file)


\begin{multicols}{2}

% Sections

\section{1. INTRODUCTION}

The study of the pre-main sequence (PMS) stars is important to learn more about the early stage of stellar evolution. 
The most important characteristic of PMS stars is the photometric and spectroscopic variability, discovered at the beginning of their study. 

In the fields of star formation the PMS stars form in groups called stellar associations or stellar aggregates. 
While the high-mass star formation takes place generally in the big molecular clouds, the low-mass stars can be formed even in the smallest regions.
Many PMS stars are found to be members of photometric and spectroscopic double or multiple systems (see Herbig 1962, Mathieu et al. 1989, Reipurth $\&$ Zinnecker 1993, Reipurth et al. 2002).
The PMS stars are rare among field stars, because the stars spend less then 1$\%$ of his life during the PMS evolutionary stage.
PMS stars are separated into two types $-$ low mass (M $\leq$ 2M$_{\odot}$) T Tauri stars (TTS) and the more massive (2M$_{\odot}$ $\leq$ M $\leq$ 8M$_{\odot}$) Herbig Ae/Be stars (HAEBES). 

The study of TTS began in the mid-20th century with the pioneering work of Joy (1945). 
TTS exhibit irregular photometric variability and emission spectra. 
They are separated into two subclasses: classical T Tauri stars (CTTS), surrounded by spacious accreting circumstellar disks, and weak-line or naked T Tauri stars (WTTS) without evidence of disks accretion (M\'{e}nard \& Bertout 1999). 
CTTSs are distinguished from WTTSs by their strong H$\alpha$ emission line and by significant infrared and ultraviolet excesses. 
Usually, the upper limit of EW(H$\alpha$)$\leq$5$\AA$ is used for defining the WTTS, but optically veiled and nonoptically veiled PMS
stars can be distinguished based on EW(H$\alpha$) only taking into account their spectral type (see White \& Basri 2003). 
Both subclasses of TTS show strong brightness variations over comparatively short time intervals (days or weeks).

According to Herbst et al. (1994, 2007) the reasons for the observed brightness variations of the PMS stars are diverse. 
The variability of CTTS is caused by a superposition of cool and hot spots on the stellar surface. 
The observed non-periodic brightness variations are produced by highly variable accretion from the circumstellar disk onto the stellar surface. 
Large amplitudes reaching up to 2-3 mag in $V$-band are observed in CTTS. 

The variability of WTTS, is due to the rotation of the stellar surface covered by cool spots or groups of spots. 
Some TTS shows clear expressed periodicity (see Herbst et al. 2007; Scholz et al. 2011). 
The periods of variability in WTTS are registered on time scales of days and with amplitudes reaching 0.8 mag. in $V$-band. 
The brightness variability in WTTS can be due also to flare-like variations in $B$- and $U$-band. 
Flares are random with varying duration and amplitudes of brightness and without periodicity (Herbst et al. 2007).

Drops in brightness, lasting from few days to several weeks or months, are observed in the early type of CTTS and HAEBES. 
Usually the decreases in brightness are non-periodic with amplitudes reaching 2.8 mag. in $V$-band. 
Presumably they are result of circumstellar dust or clouds obscuration (Grinin et al. 1991; Herbst et al. 2007). 
The prototype of this group of PMS stars with intermediate mass, named UXors, is UX Orionis. 
In very deep minima the color indexes of UXors commonly becomes bluer (color reverse), or so called ''blueing effect'' (see Bibo \& Th\'{e} 1990).

Some of CTTS undergo eclipses with irregular depths and periodicity for several days, probably caused by warped inner disk viewed near to edge-on (Bouvier et al. 1999, 2007). The prototype of this group of PMS stars is AA Tau, and the number of these objects has steadily increased during the recent years. (Barsunova et al. 2016, Sousa et al. 2016, Rodriguez et al. 2017, Schneider et al. 2018).
As the warp in the accretion disc is presumably caused by the magnetic field of the star, the recurrence period of these eclipses, equal to a few days, is actually tracing the period of the Kepler rotation of the gas disk at a distance of the corotation radius from the star.

Our photometric and spectral study of PMS stars has been done in the dark clouds at vicinity of the reflection nebula NGC 7129. 
The region is immersed in a very active and complex molecular cloud (Hartigan \& Lada 1985; Miranda et al. 1993). NGC 7129 is apparently a part of a larger structure, called the Cepheus Bubble (Kun et al. 1987) and its represents a region with active star formation (Magakian \& Movsessian 1997; Kun et al. 2008, 2009). A large number of TTS, HAEBES, Herbig-Haro objects, collimated jets and cometary nebula are observed in this region.
The distance to NGC 7129, is determined as 1.26 kpc by Shevchenko \& Yakubov (1989), as 1.15 kpc by Straiz\v{y}s et al. (2014) and as 800 pc by {\'A}brah{\'a}m et al. (2000). 
The age of the NGC 7129 star forming region determined by Straiz\'{y}s et al. (2014) is 3 Myr.

The stars V391 Cep and V1 were discovered as strong H$\alpha$ emission sources in the study of Semkov \& Tsvetkov (1986). 
The variability of the star V2 was registered by Semkov (2003b) and the variability of V3 by Semkov (2000). 
According to Semkov (2003b) the stars V1, V2 and V3 are irregular variables, which belong to the class of PMS stars. 
The variability in the star V4 is discovered during the present study. 

The present paper is a part of our photometric and spectroscopic study of the PMS stars in the vicinity of NGC 7129. 
The results from our previous studies have been published in Semkov (1993a, 1993b, 1993c, 1996, 2000, 2002, 2003a, 2003b, 2004a, 2004b), Semkov et al. (1999), Ibryamov et al. (2014), Ibryamov et al. (2017).

\section{2. OBSERVATIONS AND DATA REDUCTION}

\subsection{2.1 Photometric observations}

The photometric $UBVRI$ data presented in this paper were collected in the period from February 1998 to November 2016. 
The CCD observations were obtained with four telescopes, in two observatories $-$ the 2-m Ritchey-Chr\'{e}tien-Coud\'{e} (RCC), the 50/70-cm Schmidt and the 60-cm Cassegrain telescopes of the Rozhen National Astronomical Observatory in Bulgaria and the 1.3-m Ritchey-Chr\'{e}tien (RC) telescope of the Skinakas Observatory\footnote{Skinakas Observatory is a collaborative project of the University of Crete, the Foundation for Research and Technology, Greece, and the Max-Planck-Institut f{\"u}r Extraterrestrische Physik, Germany.} of the University of Crete in Greece. The total number of the nights used for observations is 295.
The observations were performed with eight different types of CCD cameras. Their technical parameters and specifications are given in Ibryamov et al. (2015).

All frames were taken through a standard Johnson$-$Cousins $UBVRI$ set of filters. The frames obtained with the CCD cameras at the 2-m RCC and the 1.3-m RC telescopes are bias frame subtracted and flat field corrected. All frames obtained with the CCD cameras at the 50/70-cm Schmidt and the 60-cm Cassegrain telescopes are dark frame subtracted and flat field corrected.

The photometric data were reduced using \textsc{idl} based $DAOPHOT$ subroutine. As a reference, the $UBVRI$ comparison sequence reported in Semkov (2003a) was used. All data were analyzed using the same aperture, which was chosen as 6 arc seconds radius, while the background annulus was taken from 10 to 15 arc seconds.

\vskip.5cm
\centerline{\includegraphics[width=0.9\columnwidth,
keepaspectratio]{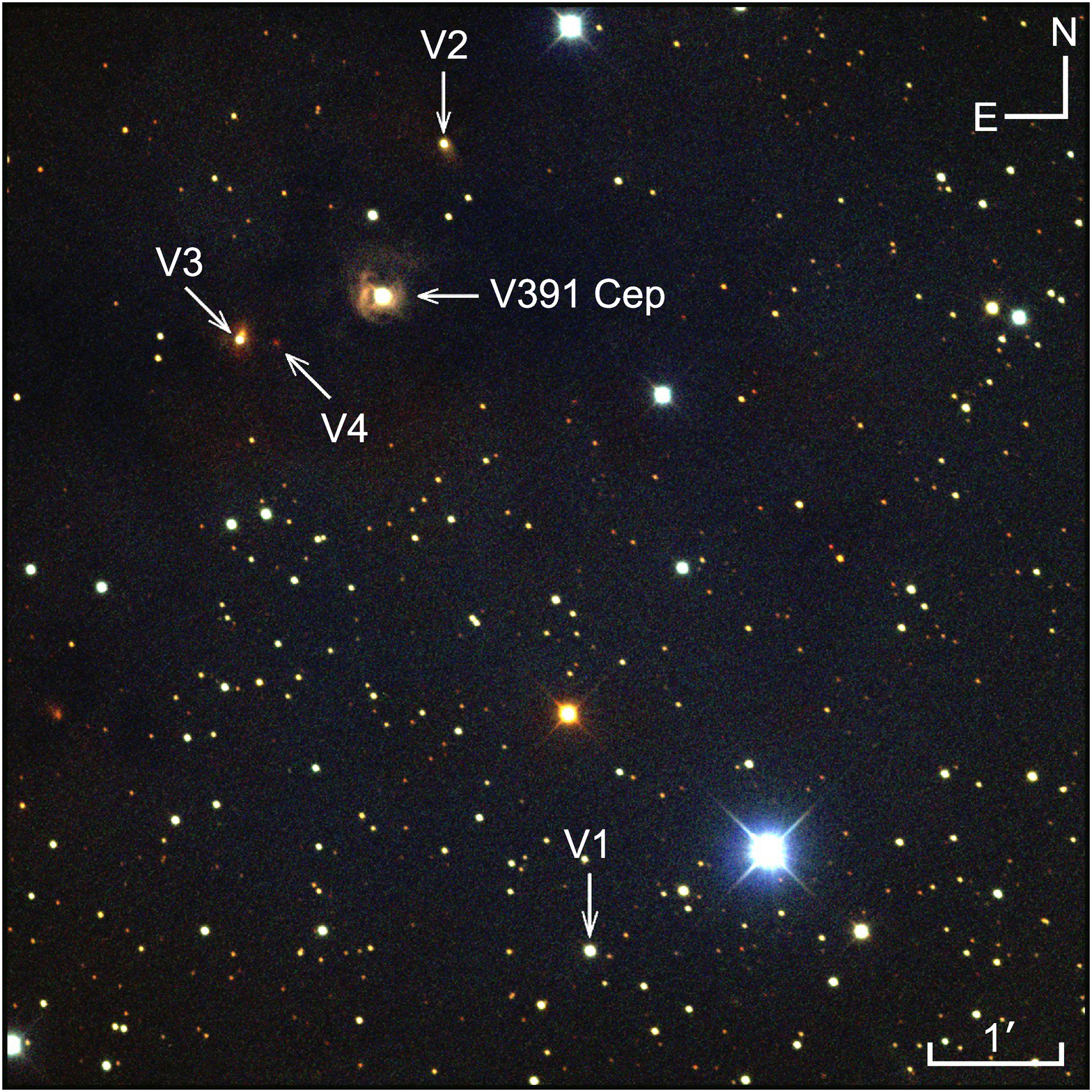}}

\figurecaption{1.}{A three-color image of the field in the vicinity of NGC 7129.}

\subsection{2.2 Spectroscopic observations}

At the time of our photometric monitoring, a total of thirteen medium-resolution optical spectra of the PMS stars in the region were obtained.
The spectral observations are performed in Skinakas Observatory with the focal reducer of the 1.3-m RC telescope and ISA 608 spectral CCD camera ($2000\times800$ pixels, 15$\times15$ $\mu$m).
Two grating 1300 lines/mm and 600 lines/mm and a 160 $\mu$m slit were used.
The combination of gratings and slit yield a resolving power $\lambda/\Delta\lambda$ $\sim$ 1500 at H$\alpha$ line for the 1300 lines/mm grating and $\lambda/\Delta\lambda$ $\sim$ 1300 for the 600 lines/mm grating.
The exposures of the objects were followed immediately by an exposure of a FeHeNeAr comparison lamp and exposure of a spectrophotometric standard star.
All data reduction was performed within \textsc{iraf}. 

\section{3. RESULTS AND DISCUSSION}

Fig. 1 shows a three-color image of the field located north-east from the bright nebula NGC 7129, with the marked positions of the stars from our study. 
The three-color image was obtained on July 23, 2007 with the 1.3-m RC telescope of Skinakas Observatory. On the figure small cometary nebulae around V391, V2 and V3 are clearly seen.

Using $JHK_{s}$ 2MASS magnitudes of the stars from our study we plot the two-color diagram to identify stars with infrared excess, indicating the presence of a disk. Fig. 2 shows the location of main sequence (brown line) and giant stars (orange line) from Bessell \& Brett (1988), the CTTSs location from Meyer et al. (1997). A correction to the 2MASS photometric system was performed following the prescription of Carpenter (2001). The three parallel dotted lines show the direction of the interstellar reddening vectors determined for the NGC 7129 region by Straiz\v{y}s et al. (2014).

\vskip.5cm
	 \centerline{\includegraphics[width=0.9\columnwidth,
keepaspectratio]{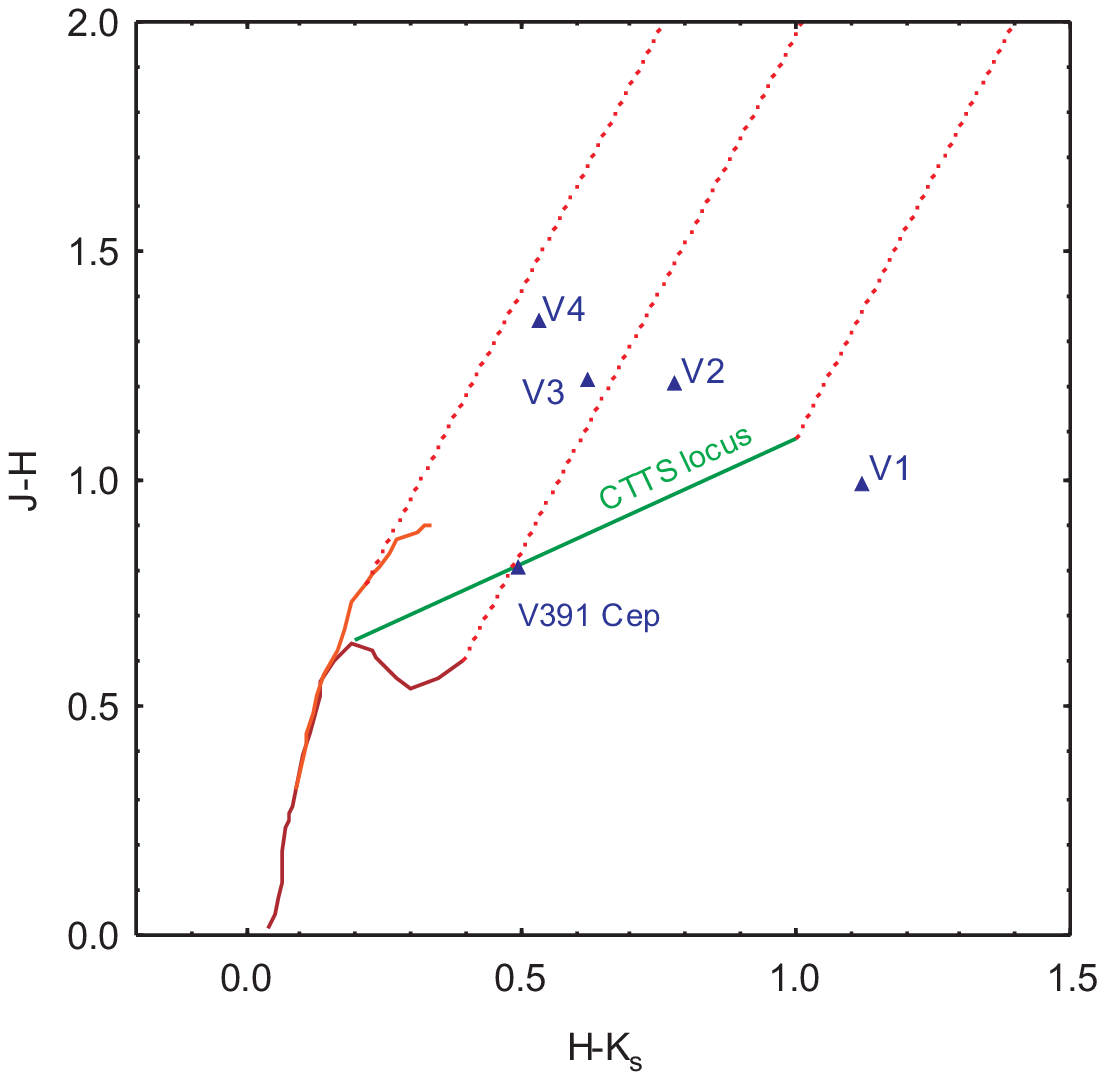}} 
	
   \figurecaption{2.}{The $J-H$ versus $H-K_{s}$ diagram for the stars from our study. The brown line denote location of main sequence, the orange line - location of giant stars, the green line - CTTS location. The dotted lines denote the direction of the interstellar reddening vectors. Blue triangles designate the PMS stars from our study.}

The positions of the stars V2, V3 and V4 indicate for a strong infrared excess, the star V391 Cep is located on the line of CTTS locus, and V1 shows colors typical of more evolved young star with limited circumstellar material. 
However, it will be shown in the following sections, that its SED, the nature of the photometric variability, and age fully correspond to the similar parameters of the remaining studied objects.
Because the stars exhibit photometric variability in all passbands, their positions in Fig. 2 can vary during the different time periods.

The optical and near-infrared parts of the SEDs of the stars from our study were constructed using all our $UBVRI$ photometric data, corrected for the interstellar extinction (Fig. 3). 
We use the extinction law given in Cardelli et al. (1989) and on the basis of visual extinction A$_{V}$ for the stars by Kun et al. (2009). 
The fluxes corresponding to zero magnitude were obtained from Bessell (1979) for the $UBVRI$ bands, and from the 2MASS All Sky Data release Web-document\footnote{http://www.ipac.caltech.edu/2mass/releases/allsky/doc/sec6${\_}$4a.html} 
for the $JHK_{s}$ bands. Photospheric SEDs have been dawn by green lines. Their values were determined from dereddened $I$ magnitudes and from the color indices corresponding to the spectral types. For comparison, red lines show the median SED of the T Tauri stars of the Taurus star-forming region (D'Alessio et al. 1999).

\end{multicols}
 \begingroup\centering
	 \includegraphics[width=4.5cm, angle=0]{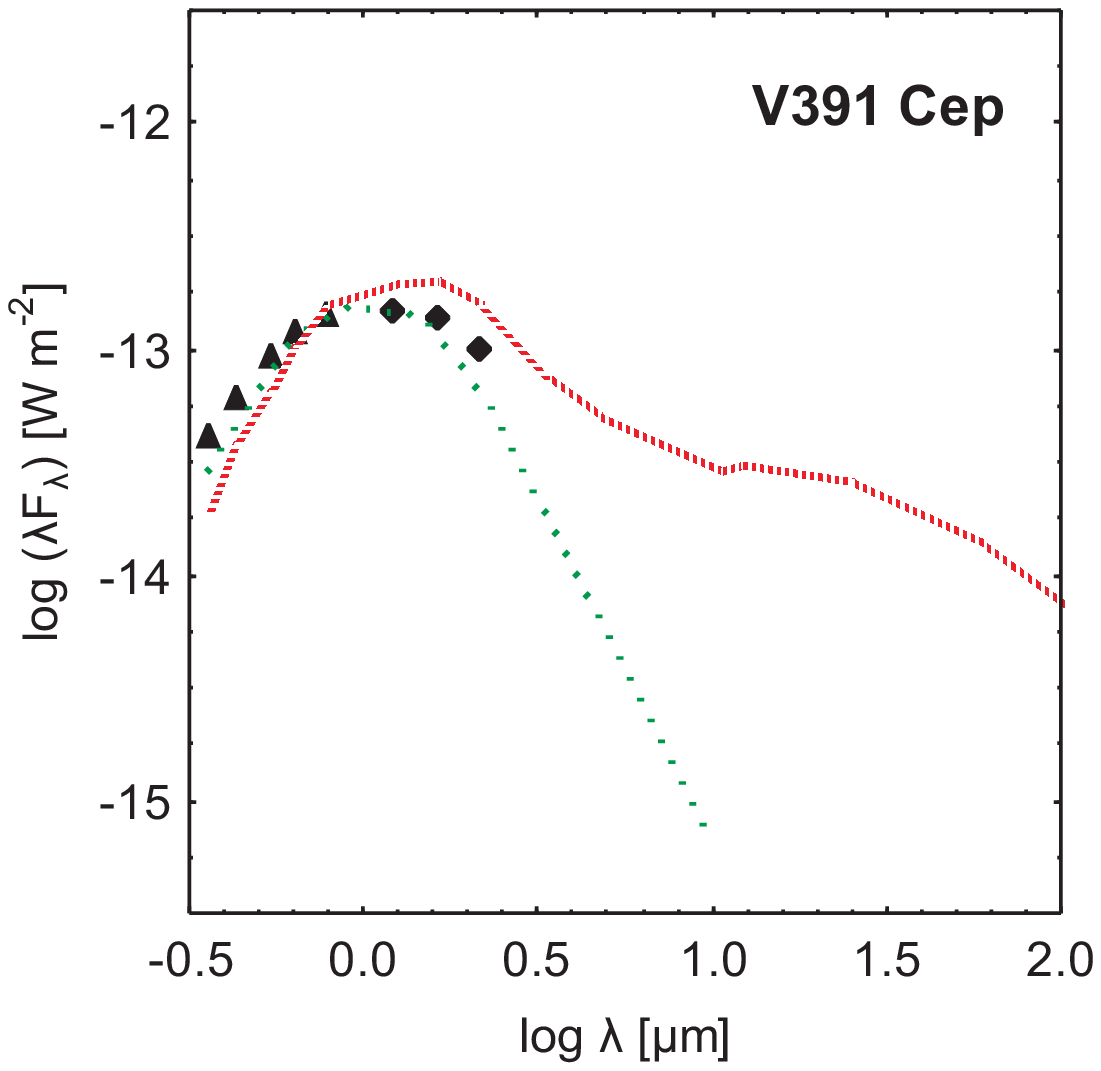}
   \includegraphics[width=4.5cm, angle=0]{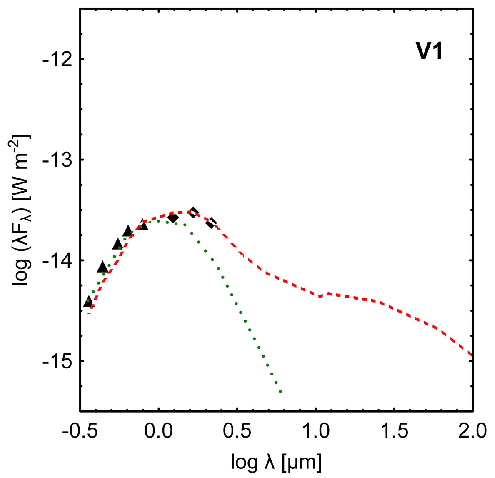}
	 \includegraphics[width=4.5cm, angle=0]{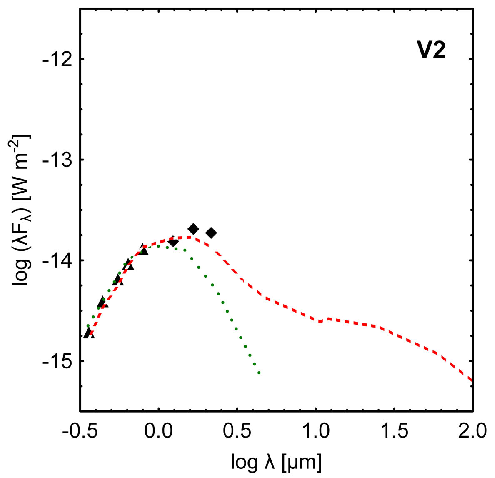}
	 \includegraphics[width=4.5cm, angle=0]{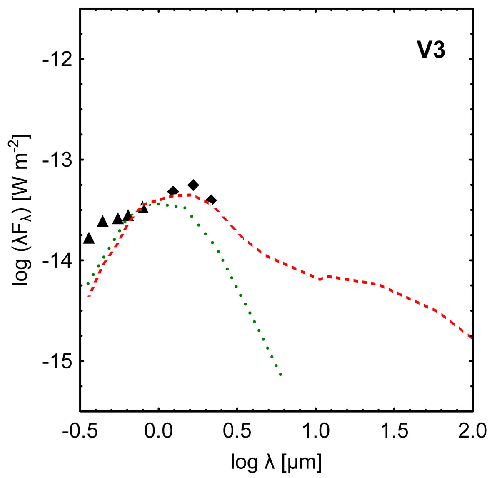}
	 \includegraphics[width=4.5cm, angle=0]{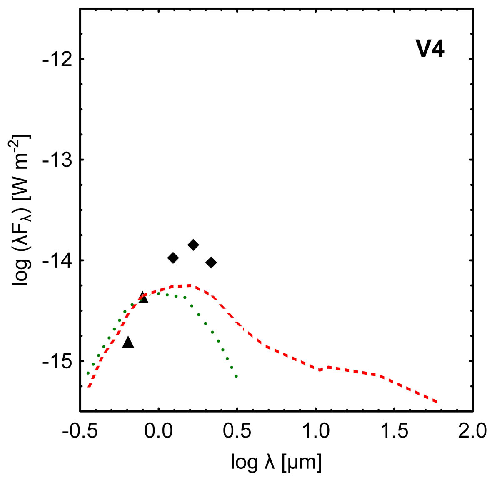}
   \figurecaption{3.}{SEDs corrected for interstellar extinction of the stars from our study, based on our $UBVRI$ photometry (triangles) and 2MASS (diamonds).}
   \endgroup
\begin{multicols}{2}

\subsection{3.1 V391 Cephei}

V391 Cep is located at about 33 arc minutes from the center of NGC 7129 and at about 28 arc minutes from another T Tau star - V350 Cep (see Ibryamov et al. 2014). 
V391 Cep was discovered as a strong H$\alpha$ emission source by Semkov $\&$ Tsvetkov (1986) and included in the list of H$\alpha$ emission stars published by Kun (1998).
Spectral observations of V391 Cep (Semkov 1993c; Kun et al. 2009) suggest that its spectrum is similar to CTTS spectrum, with strong H$\alpha$ line and presence of emission lines of oxygen, iron, magnesium and others metals. 
Kun et al. (2009) defined the spectral class of V391 Cep as K5 and determined its mass as 1.15 M$_{\odot}$, its effective temperature as 4350 K and its age as 0.2 Myr.

The CCD observations performed by Semkov (1993c) shows the presence of a small cometary nebula around the star, which is not visible on the Palomar Observatory Sky Survey prints and on the photographic plates collected in the field of NGC 7129 (Semkov 1993b).
From 1986 to 2002 the variability of brightness in $B$-band changes significantly, the amplitude decreases gradually from 2.1 mag to 0.3 mag. 
According to Semkov (2003a) V391 Cep is a CTTS and the change in photometric activity can be caused by an irregular accretion rate.

The subsequent observations performed by Semkov (2000) showed that V391 Cep is a visual double system with a deep infrared component. The distance between the two components is 3 arc seconds. Semkov (2000) determined color index $R-I$ for the second component and concluded that the object is likely a cool star in the first stages of star formation, forming double system of PMS stars with V391 Cep.

   \end{multicols}
   
   \includegraphics[width=13cm, angle=0]{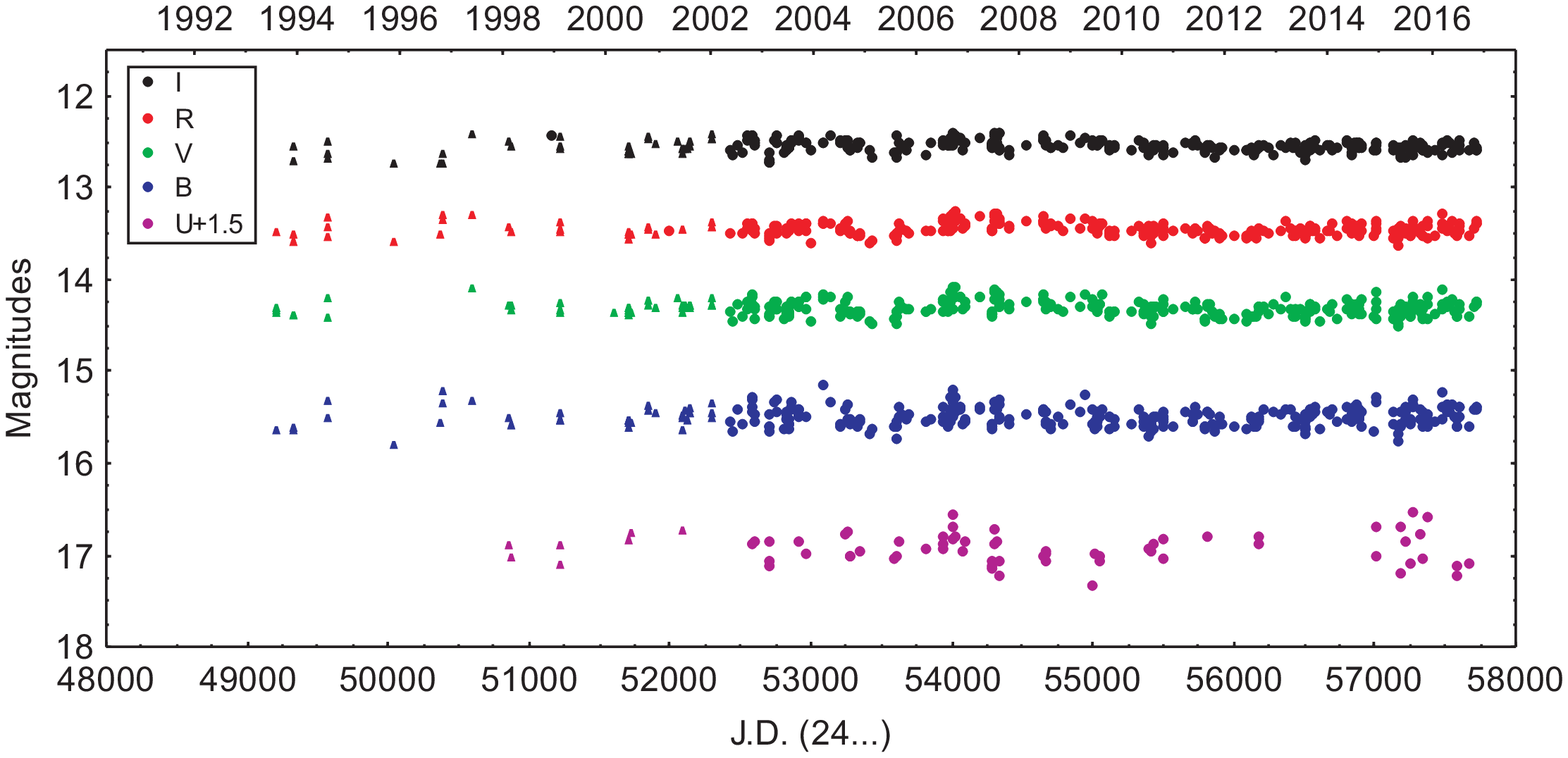}
   \figurecaption{4.}{CCD $UBVRI$ light curves of V391 Cep for the period August 1993$-$November 2016.}
   
		\begin{multicols}{2}

The results of our long-term multicolor CCD observations of V391 Cep will be accessible through the CDS database. The average value of the errors in the measured magnitudes are $0.01$-$0.02$ mag for the $I$- and $R$-band data, $0.01$-$0.03$ for the $V$-band data, $0.02$-$0.04$ for the $B$-band data, and $0.06$-$0.10$ for the $U$-band data. 

The $UBVRI$ light curves of the star from all our CCD observations (Semkov 2003a and the present paper) are shown in Fig. 4. In the figure, circles denote CCD photometric data published in the present paper and triangles $-$ CCD photometric data from Semkov (2003a).

The brightness of V391 Cep during the period of all our CCD observations 1993$-$2016 varies in the range 12.39$-$12.76 mag for the $I$-band, 13.25$-$13.63 mag for the $R$-band, 14.09$-$14.51 mag for the $V$-band, 15.15$-$15.82 mag for the $B$-band and 15.03$-$15.82 mag for the $U$-band. 
Such photometric characteristics (variability with small amplitude in a time scale of days) are typical for both WTTS and CTTS.

The long-term $UBV$ light curves of V391 Cep from all available published observations (1980$-$2016) are presented in Fig. 5. The circles denote CCD observations published in present study; triangles denote CCD photometric data from Semkov (2003a) and diamonds denote the photographic data from Semkov (1993b).

	\end{multicols}
   
   \includegraphics[width=13cm, angle=0]{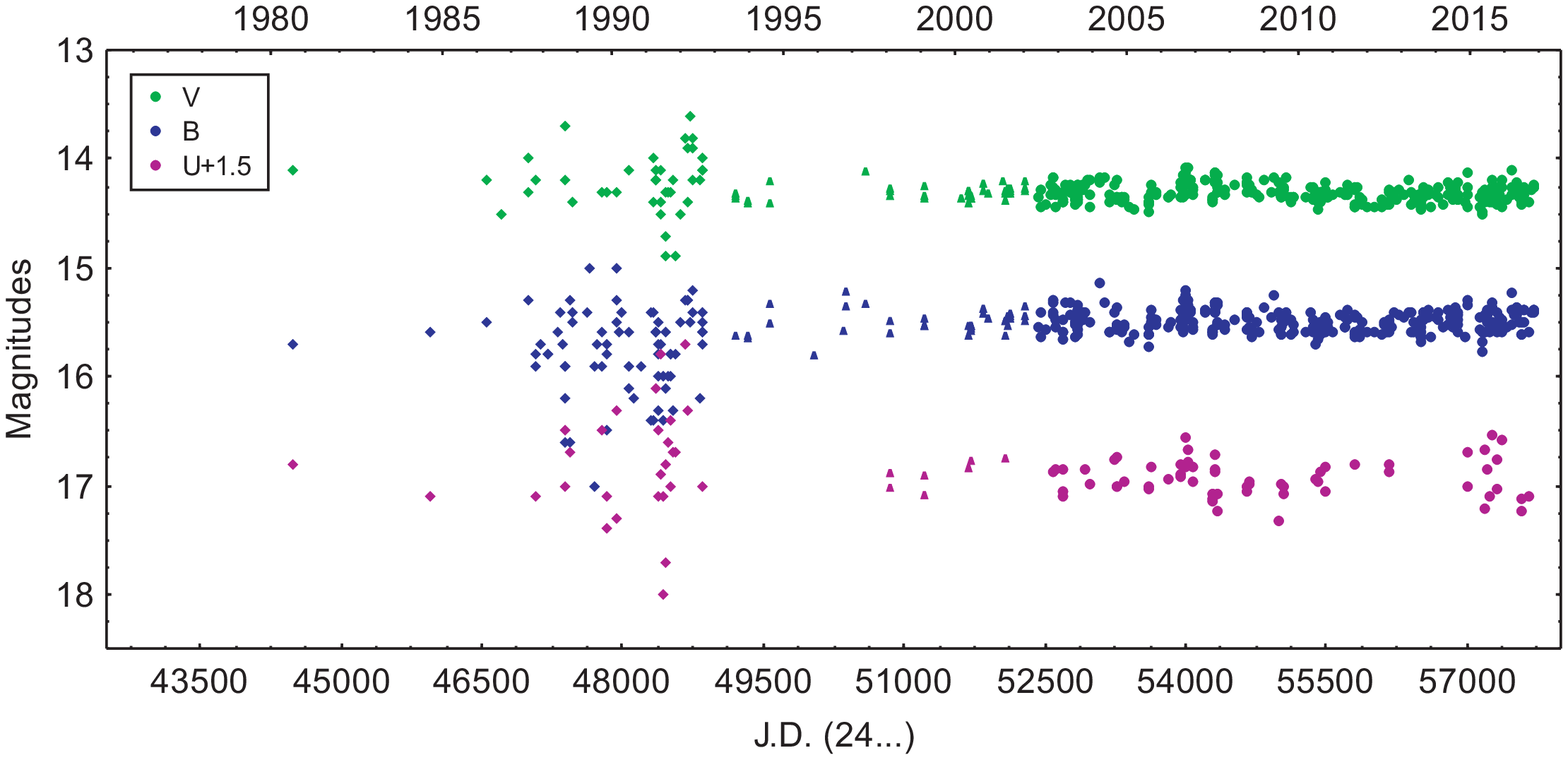}
   \figurecaption{5.}{$UBV$ light curves of V391 Cep from all available published observations.}

   \begin{multicols}{2}

The available data suggest that in the period 1986$-$1992 the star exhibit strong irregular variability. 
Increases and fading events in the star's brightness are seen during the same period. 
The reasons for observed strong photometric variability may be of a different nature. 
Variable accretion activity, existence of cool and hot spots of the stellar surface, obscuration of the star by circumstellar material are possible reasons. 
The observed amplitudes in the period 1987$-$1992 are $\Delta V$ = 1.3 mag, $\Delta B$ = 2.0 mag and $\Delta U$ = 2.3 mag. 
After 1992 the brightness of V391 Cep varies with comparatively small amplitudes around some average level. Evidences of periodicity in the brightness variability of V391 Cep are not detected.
We believe this change in the amplitude of the variability is real, since the errors from photographic observations do not exceed $\pm$0.2 mag.

The measured color indexes $V-R$ and $B-V$ versus stellar $V$ magnitude, and $U-B$ index versus $B$ magnitude for the period of all our CCD observations are plotted in Fig. 6.
In the figure the $V/V-R$ diagram indicates that V391 Cep becomes redder as it fades, but on the $V/B-V$ diagram a change in the color was not observed. Such color variations are typical for TTS with presence of cool spots or groups of spots on the stellar surface. It can be seen from Table 5, that V391 Cep shows a strong ultraviolet excess $-$ a characteristic of CTTS.
		
	 \end{multicols}
	\begingroup\centering 
	 \includegraphics[width=3.5cm, angle=0]{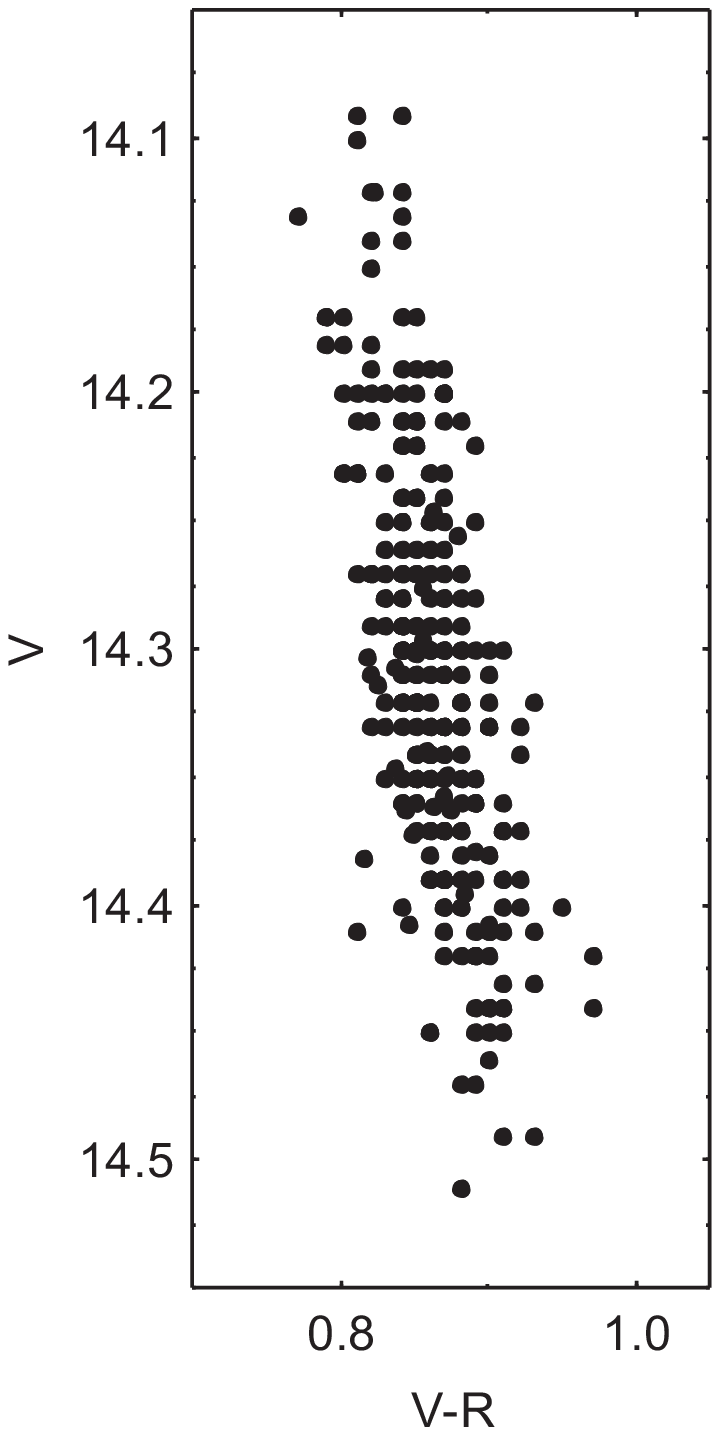}
   \includegraphics[width=3.5cm, angle=0]{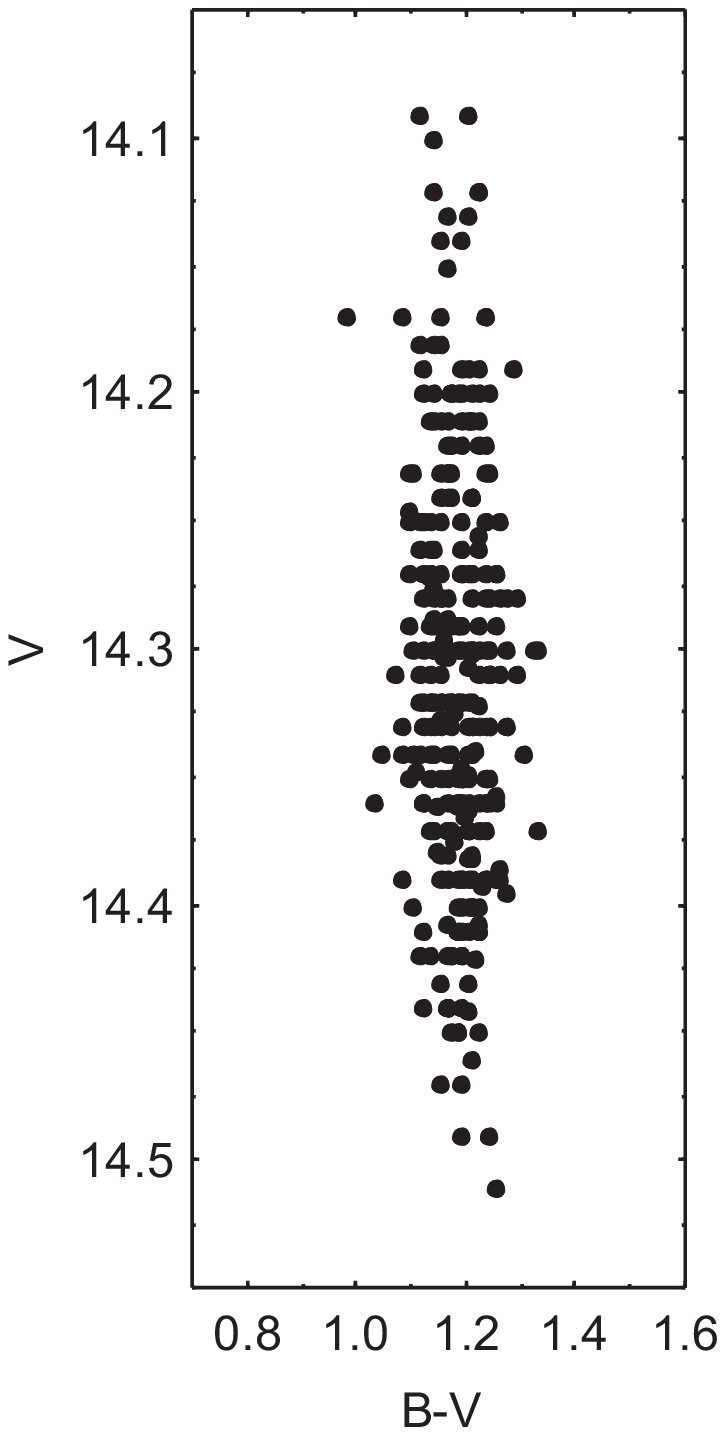}
	 \includegraphics[width=3.5cm, angle=0]{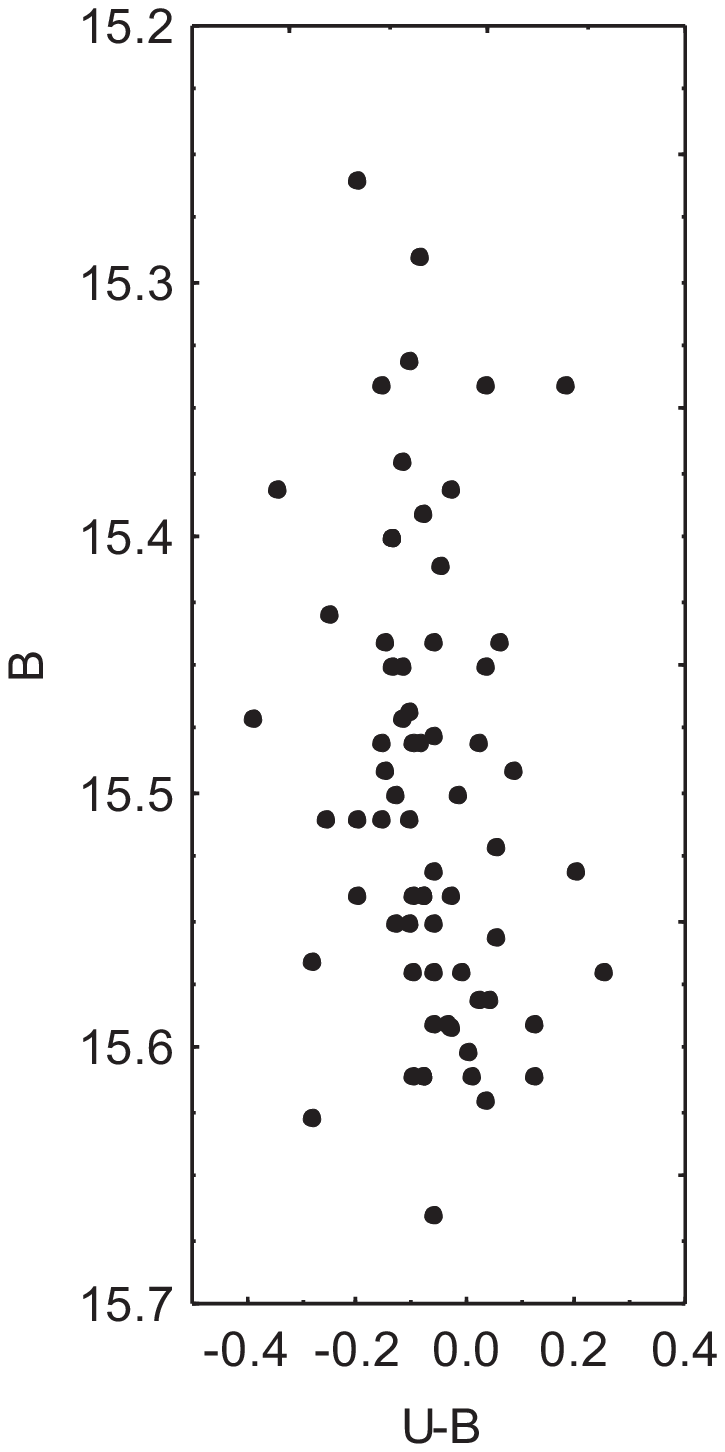}
  
	\figurecaption{6.}{The $V-R$ and $B-V$ color indexes versus $V$ magnitude, and $U-B$ index versus $B$ magnitude for V391 Cep in the period of all our CCD observations.}
   
\vskip.5cm \noindent
\parbox{\textwidth}{
{\bf Table 1. }{Main spectral lines identified in the optical spectra of V391 Cep and lines' equivalent width.}}
\begin{tabular}{llllllllllllllllll}
\hline \hline
\noalign{\smallskip}
Date&H$\delta$&H$\beta$  &Mg I &Fe I   &Fe II &[O I] &He I &NaD &[O I] &O I  &Fe II &Fe II &H$\alpha$ \\
yyyymmdd    &4102 &4861 &5173 &5270 &5316 &5577 &5876 & 5890 &6300  &6431 &6456  &6516  &6563 \\
\hline
\noalign{\smallskip}
20010907&&&&-1.2&-2.2&-9.4&-0.5&-1.4&-4.5&-0.8&-1.0&-1.1&-48.8\\
20030803&&&-7.3&-2.9&-2.6&-6.1&-1.3&-1.9&-3.3&-0.8&-0.9&&-62.9\\
20030804&&&-6.8&-2.9&-2.6&-4.2&-1.6&-2.5&-3.5&-0.6&-1.1&-1.5&-60.0 \\
20040819&-14.3&-14.3&-7.1&-4.9&-6.0&-9.8&-0.7&-1.5&-4.5&&&&-63.4\\
20050812&-8.8&-16.6&-5.9&-3.8&-3.4&-4.6&-0.9&-1.4&-9.3&&&&-88.0\\
20061004&&&&&&&-1.6&-3.0&-3.1&-1.2&-1.1&-1.3&-57.0 \\
20100824&&-8.6&-5.7&-3.0&-2.6&-3.1&-1.3&-1.8&-2.5&-0.7&-0.8&-1.1&-58.0 \\
\hline \hline
\end{tabular}
\endgroup 

\begin{multicols}{2}

In Fig. 7 we show contour plot from a CCD frame of V391 Cep and surrounding nebula. The frame is taken on July 25, 2008 with the 1.3-m RC telescope of Skinakas Observatory. 
The image was taken in relatively good atmospheric seeing (around 1 arc seconds). The levels of contour plot are selected arbitrarily.
The bright nebula around the star has not changed significantly during the period of our observations.

Our spectral observations of V391 Cep revealed an extensive emission spectrum without absorption lines (Table 1).
The emission lines of hydrogen, iron, magnesium, sodium and forbidden lines of oxygen dominate the spectrum of the star.
The most intensive is the H$\alpha$ line of hydrogen, for which in some spectra a weak absorption components from P Cyg profile is noticeable.
All emission lines show a strong variability of intensity and of full width at half-maximum.

\centerline{\includegraphics[width=0.9\columnwidth,
keepaspectratio]{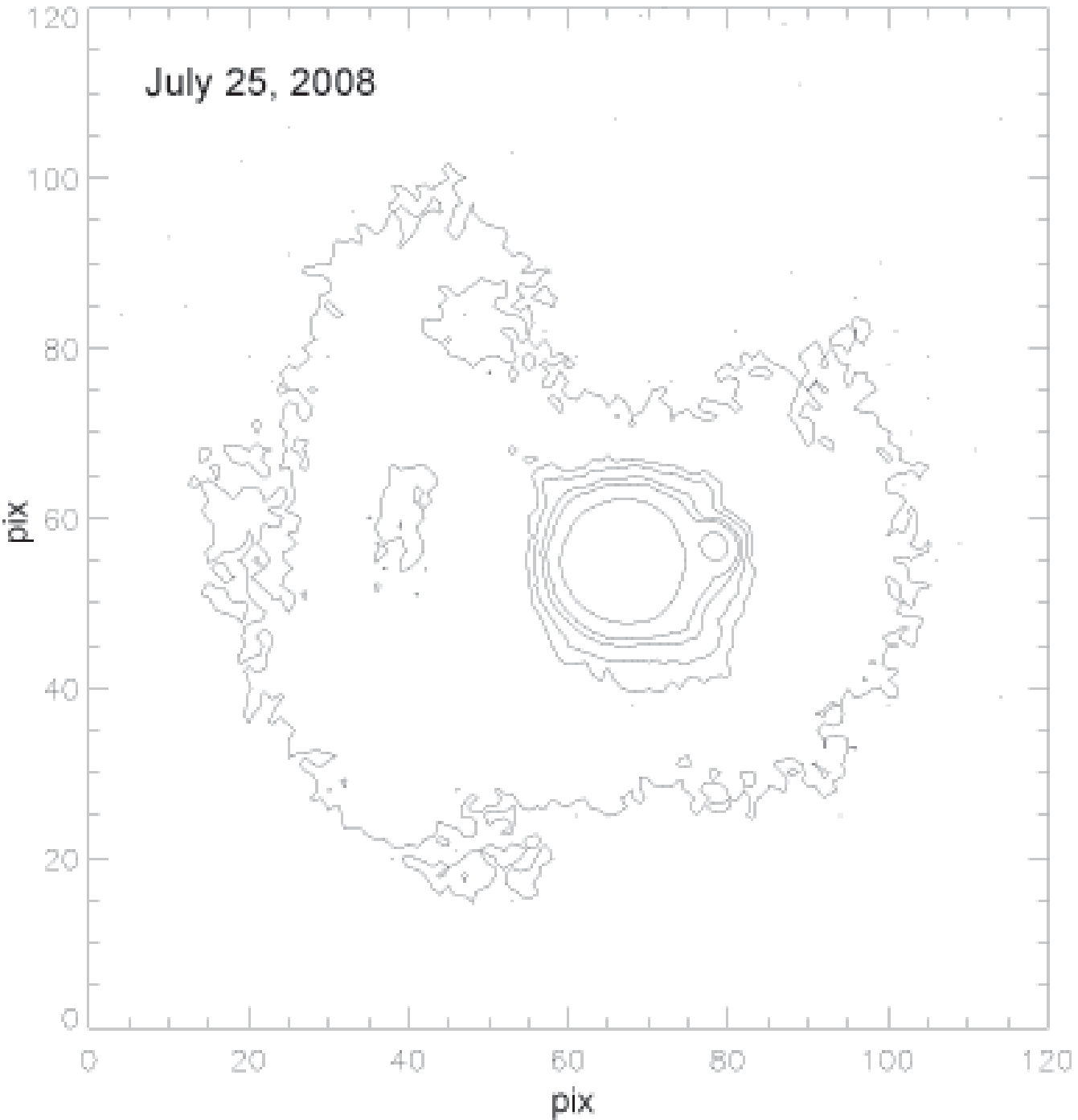}} 
	 \figurecaption{7. }{Contour plot from CCD frame of V391 Cep in $R$-band obtained with the 1.3-m RC telescope in Skinakas Observatory on July 25, 2008.}

Consequently, in the period 1987$-$1992 V391 Cep shows photometric characteristics of CTTS, but after 1992 its photometric variability is inherent of both WTTS and CTTS with a presence of hot and cool spots on the stellar surface. 
Although all our spectra were obtained in the period after 1992 clearly every spectrum is eligible for CTTS.
Our conclusion is that V391 Cep is a classical T Tauri star that undergoes periods of enhanced disk accretion alternated with periods of low accretion rates.

\subsection{3.2 V1 (2MASS J21401174+6630198, NGC 7129 S V1)}

The star 2MASS J21401174+6630198 (hereafter V1) was discovered as an H$\alpha$ emission source in the study of Semkov \& Tsvetkov (1986). The star is located at $\sim$5.4 arc minutes from V391 Cep.
Kun et al. (2009) measured $I$ = 14.34, $R$ = 15.20, $V$ = 16.18 and $B$ = 17.37 magnitudes of V1. The authors defined the spectral class of the star as K7 and determined its mass as 0.8 M$_{\odot}$, its effective temperature as 4060 K, and its age as 2.5 Myr. The spectrum of V1 contains emission lines of H$\alpha$ and HeI 6678 lines and it is classified as a CTTS spectrum.

   \end{multicols}
   \includegraphics[width=12cm, angle=0]{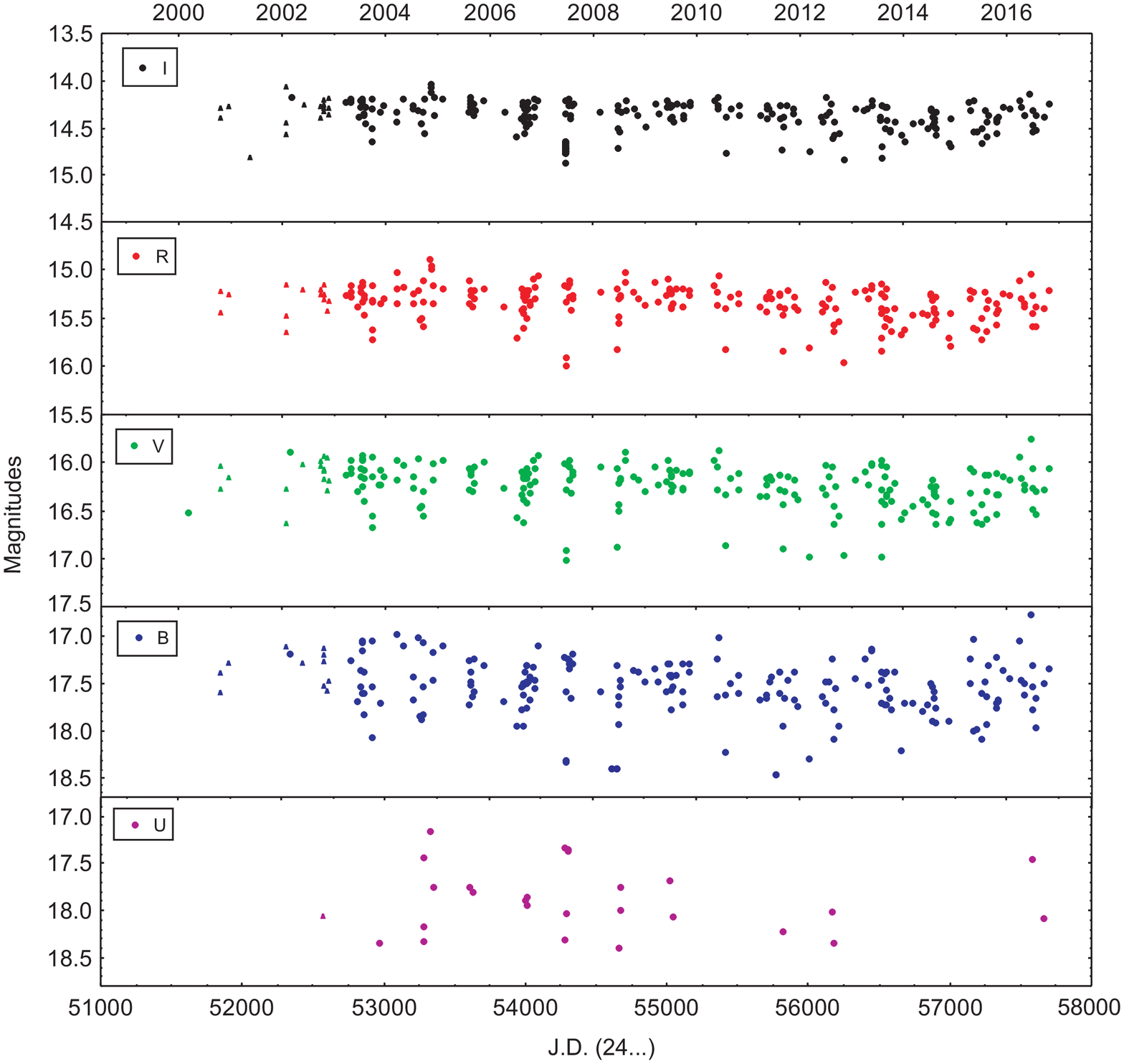}
   \figurecaption{8. }{CCD $UBVRI$ light curves of V1 for the period March 2000$-$November 2016.}
   
   \begin{multicols}{2}

Fig. 8 present $UBVRI$ light curves of V1 from all our CCD observations, obtained in the period 2000$-$2016 (Semkov 2003b and the present paper). 
In the figure, circles denote CCD photometric data published in the present study and triangles denote photometric data from Semkov (2003b). The results from our long-term multicolor CCD observations of V1 will be accessible through the CDS database. The average value of the errors in the measured magnitudes are $0.01$-$0.02$ for the $I$- and $R$-band data, $0.01$-$0.03$ for the $V$-band data, $0.02$-$0.06$ for the $B$-band data, and $0.07$-$0.13$ for the $U$-band data. 

It is seen from Fig. 8 that V1 spends most of the time at maximal light. During our photometric monitoring several declines in brightness of the star in all bands are registered. The periods of decline in the brightness are relatively short and sometimes we have only one photometric point in the minimum light.
The brightness of V1 during the period of our observations varies in the range 14.04$-$14.88 mag for the $I$-band, 14.89$-$16.01 mag for the $R$-band, 15.76$-$17.02 mag for the $V$-band, 16.79$-$18.46 mag for the $B$-band and 17.16$-$18.39 mag for the $U$-band. 
Evidences of periodicity in the brightness variability of V1 are not detected.

The measured color indexes $V-I$, $V-R$ and $B-V$ versus the stellar $V$ magnitude during the period of all our CCD observations are plotted in Fig. 9. From the figure it is seen that the star becomes redder as it fades and real color reverse is observed on the $V/B-V$ diagram. 

	\end{multicols}
   \begingroup\centering 
   \includegraphics[width=3.5cm, angle=0]{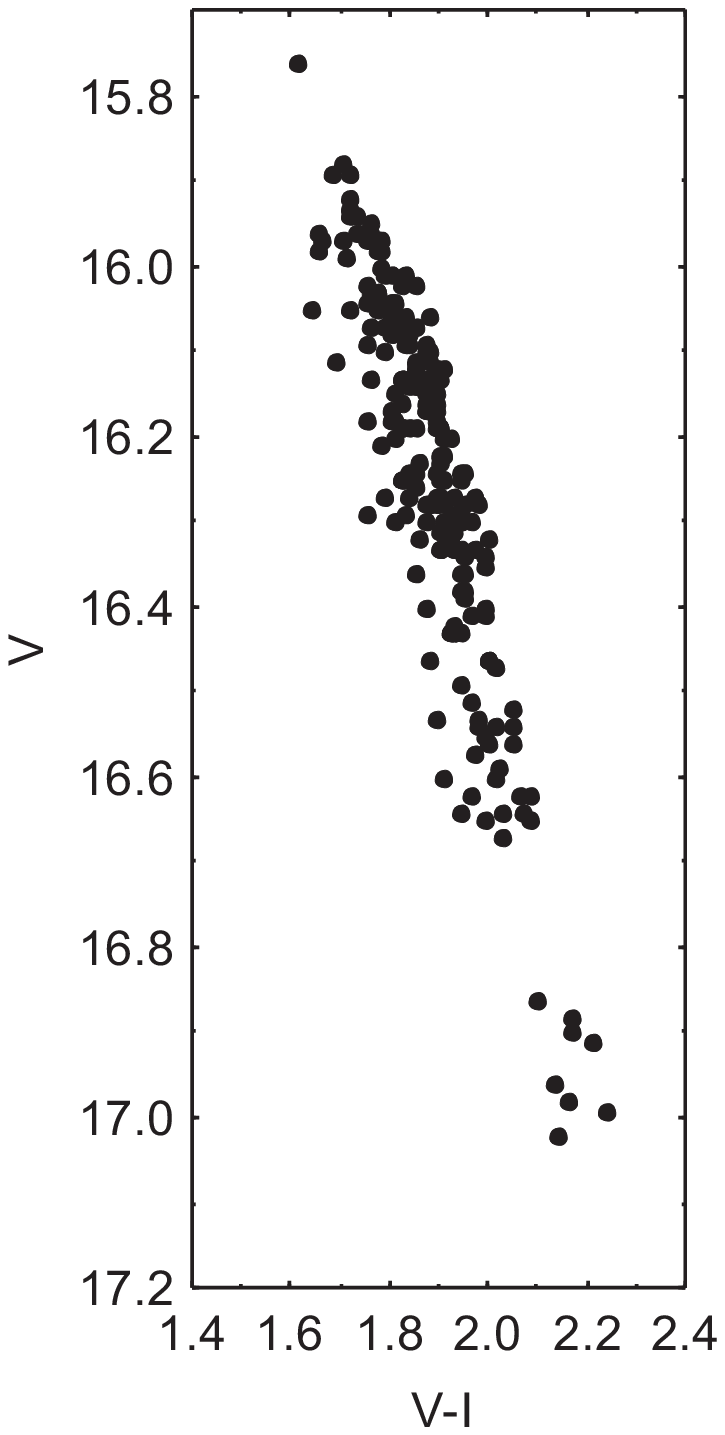}
	 \includegraphics[width=3.5cm, angle=0]{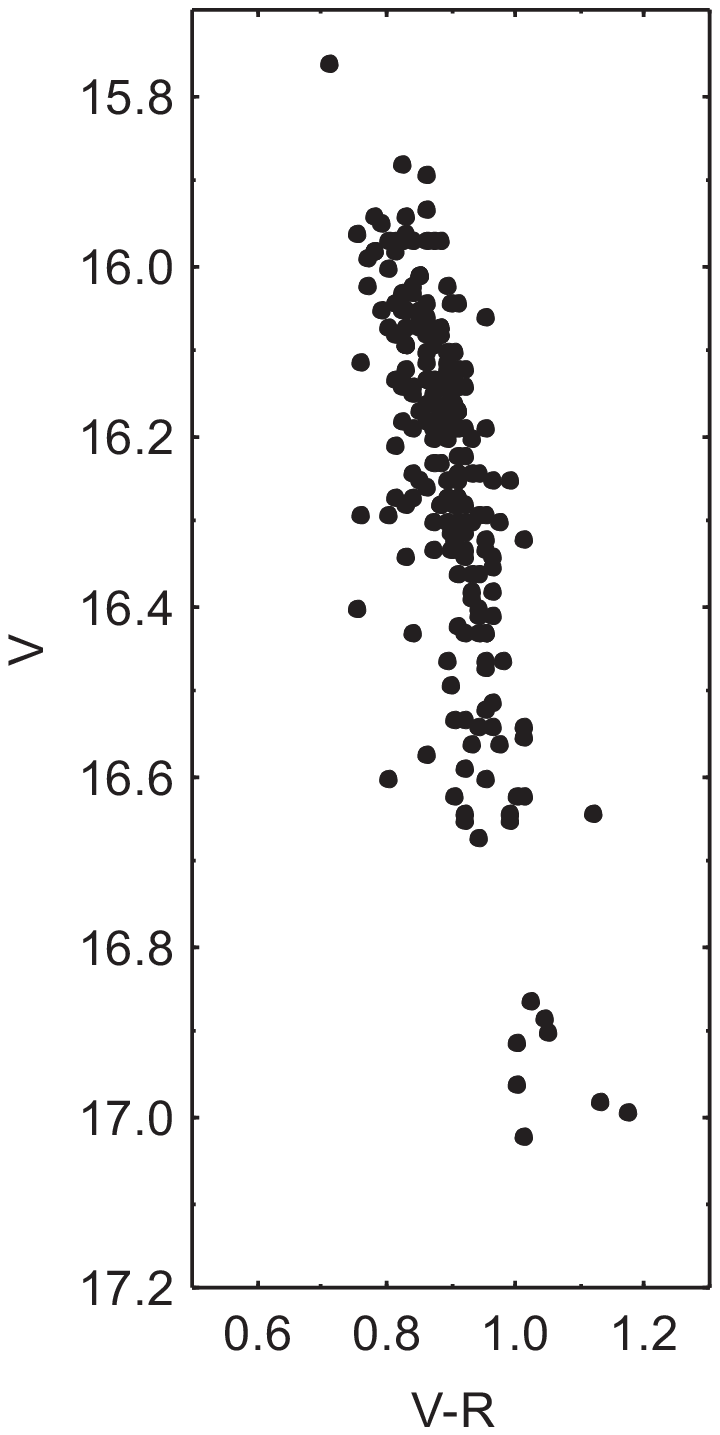}
   \includegraphics[width=3.5cm, angle=0]{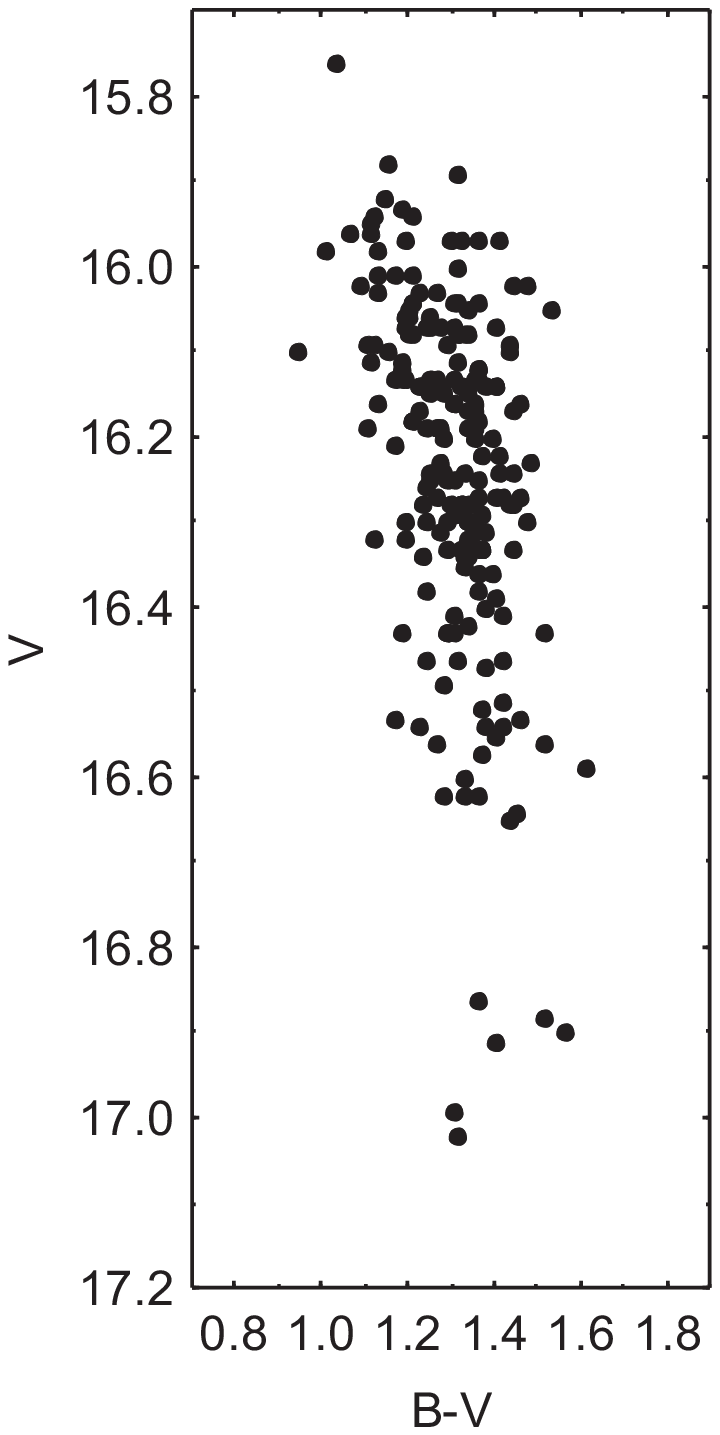}
  	\figurecaption{9. }{The $V-I$, $V-R$ and $B-V$ color indexes versus $V$ magnitude for V1 in the period of all our CCD observations.}
  \endgroup 
	
   \begin{multicols}{2}
	
The reasons of observed non-periodic drops in the brightness can be different, short irregular obscuration of the star by circumstellar dust or clouds, or eclipse from multiple bodies rotating in orbits around the star. In the case that the decreases in the brightness are caused by obscuration by circumstellar material, the observed amplitudes are too small to show indication for strong blueing effect.

The both spectra of V1 can be classified as CTTS spectra (Table 2).
The H$\alpha$ emission line and the forbidden line of oxygen (OI $\lambda$5577) are the most intensive, but lines of iron are also seen in emission.
The identified absorption lines are of the sodium doublet and lithium (LiI $\lambda$6707).
Unlike the other stars from our study the spectra of V1 showed no significant spectral variability.

Therefore, during the periods outside of the deep minima V1 can be classified as CTTS with weak activity.
The observed drops in brightness probably are caused by clouds of dust orbiting the star, but their size and density are relatively small to cause a classical UXor phenomenon. 

\end{multicols}
\vskip.5cm \noindent
\parbox{\textwidth}{
{\bf Table 2. }{Spectral lines identified in the optical spectra of V1 and lines' equivalent width.}}
\begin{tabular}{cllllllll}
\hline \hline
\noalign{\smallskip}
Date    &Fe I &[O I] &NaD&[O I] &O I &H$\alpha$&Li I &Fe I \\
YYYYMMDD&5461 &5577 &5890,96 &6300 &6431 &6563 &6707 &6829 \\
\hline
\noalign{\smallskip}
20010908&-2.5&-25.3&&-3.4&-0.6&-20.7&0.5&-1.1 \\
20030803&-1.4&-21.3&3.1&-1.5&&-19.5&0.5&-1.6 \\
\hline \hline
\end{tabular}
\begin{multicols}{2}

\subsection{3.3 V2 (2MASS J21402277+6636312, NGC 7129 S V2)}

Variability of the star 2MASS J21402277+6636312 (hereafter V2) was registered by Semkov (2003b). The star is located at $\sim$1.3 arc minutes from V391 Cep.
The authors defined the spectral class of the star as M0 and determined its mass as 0.6 M$_{\odot}$, its effective temperature as 3850 K, and its age as 3.0 Myr. The spectrum of V1 contains emission line of H$\alpha$ and it is classified as a CTTS spectrum.

In the Fig. 1 the existence of a small cometary nebula around V2 can be seen. 
The $UBVRI$ light curves of the star from all our CCD observations (Semkov 2003b and the present paper) are shown in Fig. 10. 
The symbols used are as in Fig. 8. 
The results of our long-term multicolor CCD observations of V2 will be accessible through the CDS database. The average value of the errors in the measured magnitudes are $0.01$-$0.03$ for the $I$- and $R$-band data, $0.01$-$0.05$ for the $V$-band data, $0.02$-$0.07$ for the $B$-band data, and $0.07$-$0.15$ for the $U$-band data. 

   \end{multicols}
     \includegraphics[width=12cm, angle=0]{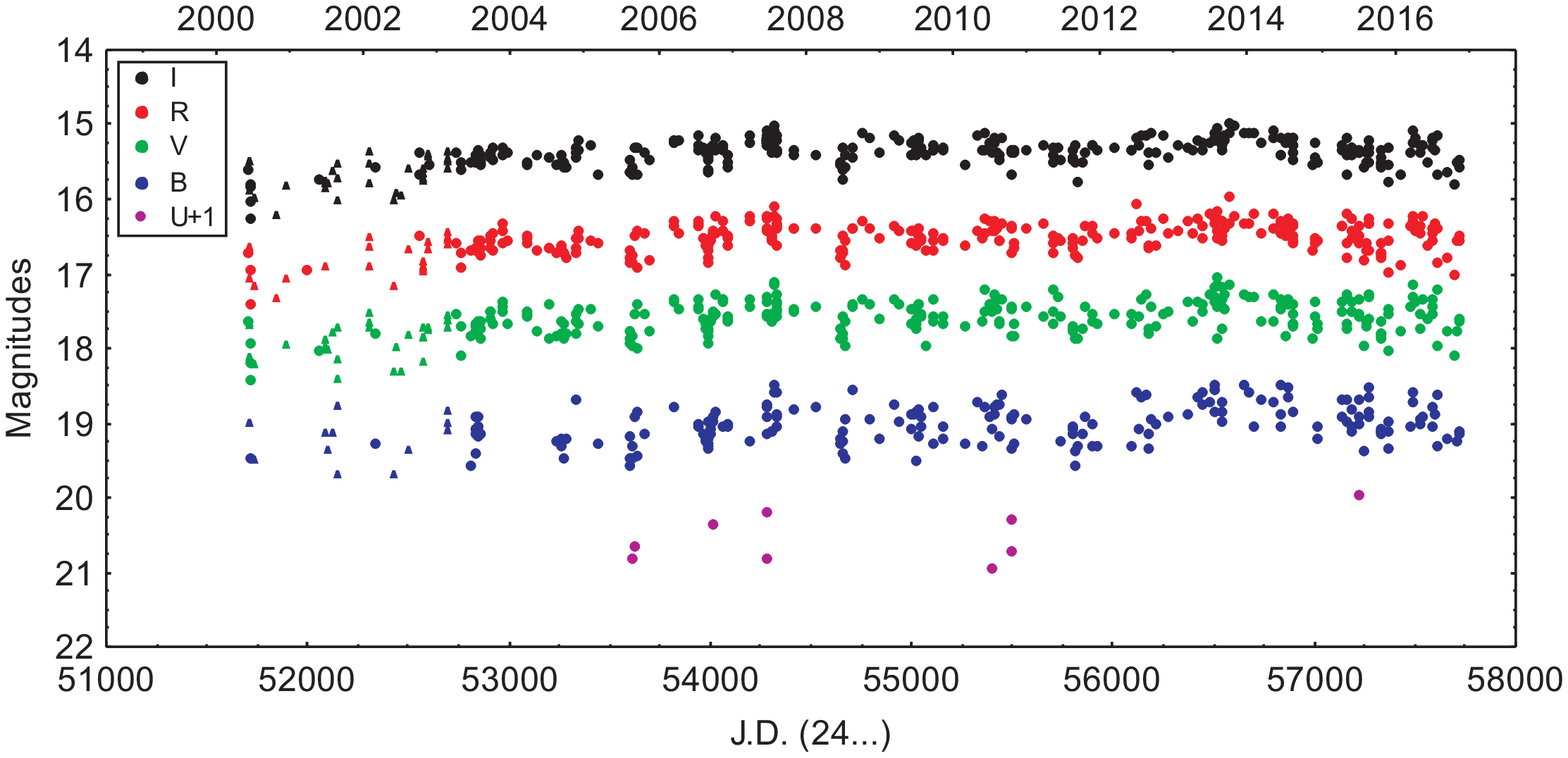}
   \figurecaption{10. }{CCD $UBVRI$ light curves of V2 for the period June 2000$-$November 2016.}
      \begin{multicols}{2}
	
From Fig. 10 it can be seen that during the period of our observations V2 shows variability in all bands. This variability includes short time rises and decreases of the star's brightness with small amplitudes.
The brightness of the star during the period of our observations varies in the range 14.97$-$16.27 mag for the $I$-band, 15.98$-$17.41 mag for the $R$-band, 17.04$-$18.42 mag for the $V$-band, 18.48$-$19.69 mag for the $B$-band and 18.98$-$19.94 mag for the $U$-band. Because the limit of our photometric data we have only nine photometric points in the $U$-band.

The measured color indexes $V-I$, $V-R$ and $B-V$ versus stellar $V$ magnitude for the period of our CCD observations are plotted in Fig. 11. 

	 \end{multicols}
   \begingroup\centering
   \includegraphics[width=3.5cm, angle=0]{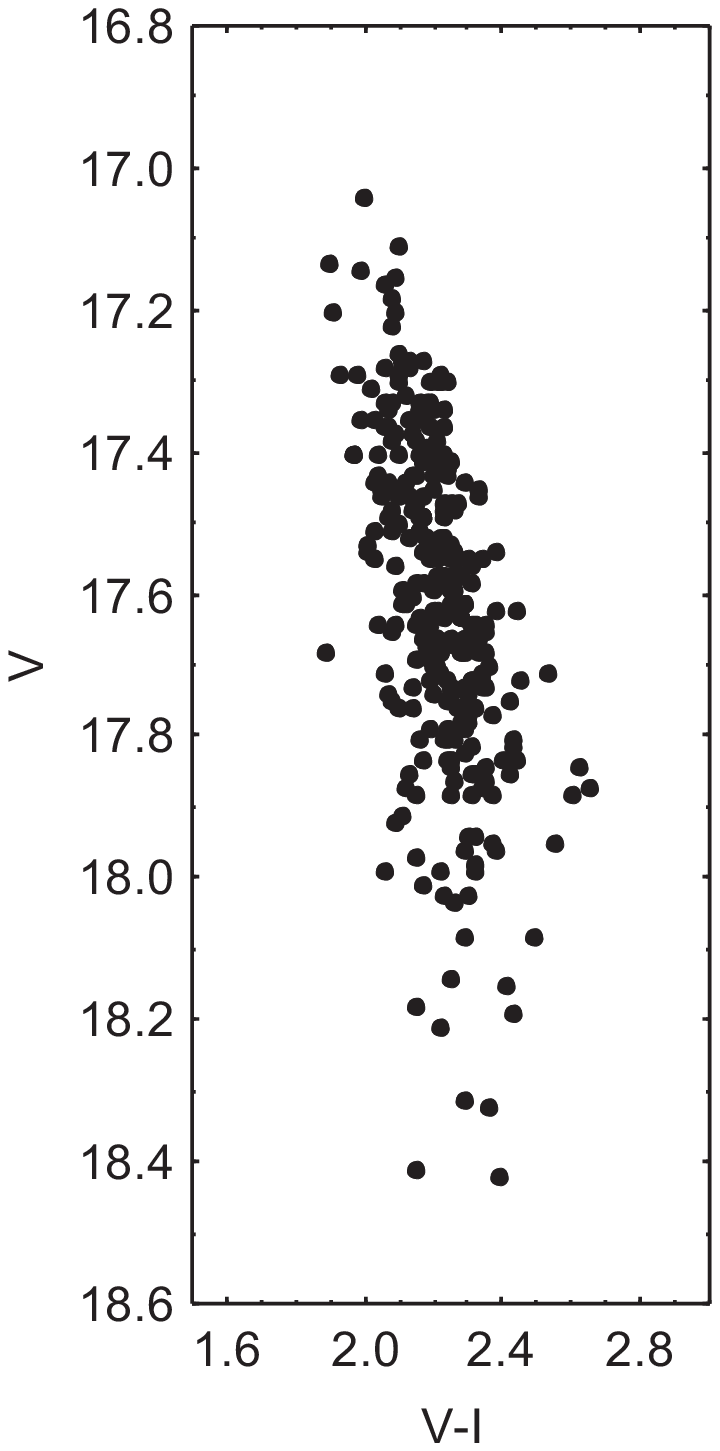}
	 \includegraphics[width=3.5cm, angle=0]{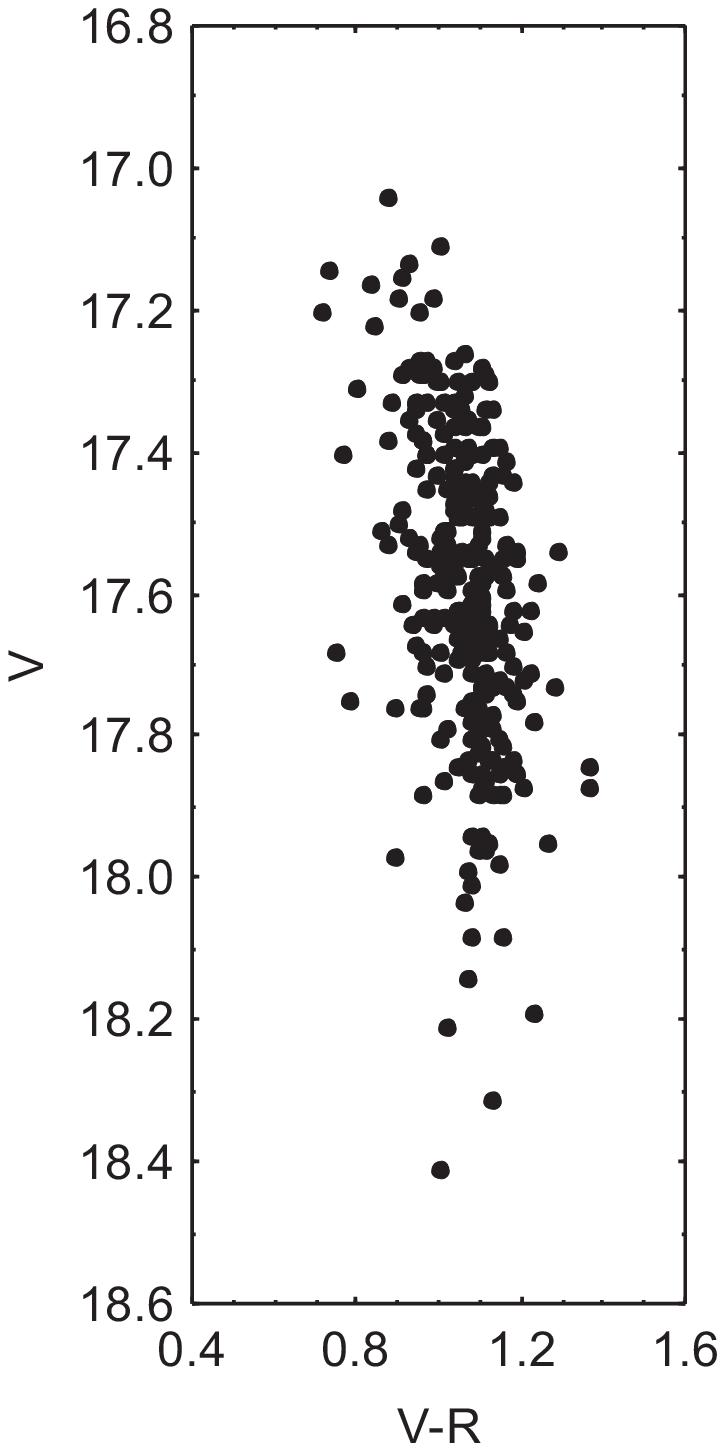}
   \includegraphics[width=3.5cm, angle=0]{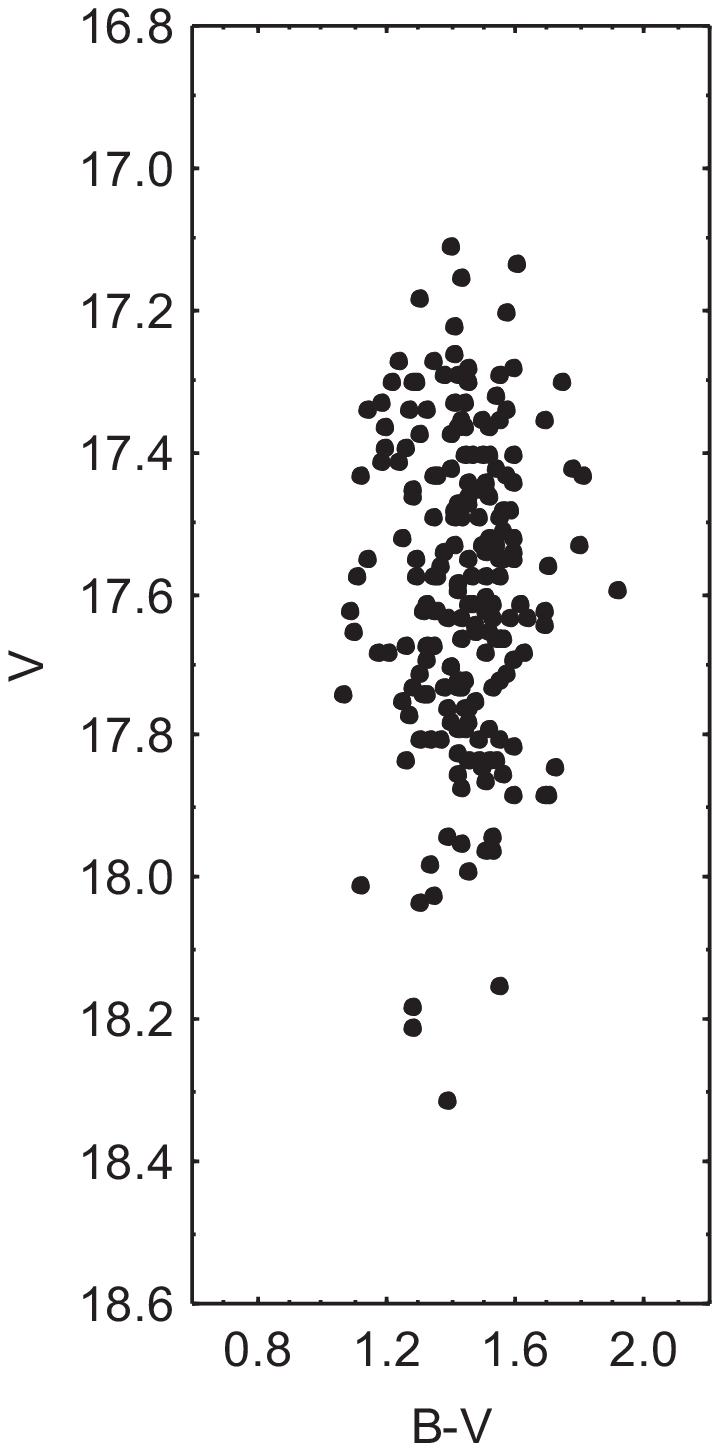}
   \figurecaption{11. }{The $V-I$, $V-R$ and $B-V$ color indexes versus $V$ magnitude for V2 in the period of all our CCD observations.}

\vskip.5cm \noindent
\parbox{\textwidth}{
{\bf Table 3. }{Spectral lines identified in the optical spectra of V2 and lines' equivalent width.}}
\begin{tabular}{lllllllllllllllll}
\hline \hline
\noalign{\smallskip}
Date&Fe I &[O I] &NaD&[O I] &H$\alpha$&Fe I &Fe II &Fe II &Fe I &Fe I &He I  &Fe II &Si III \\
yyyymmdd&5461  &5577 & 5890  &6300 &6563 &6829  & 6862 &6922 &6948  &7248 &7281 &7340  & 7369\\
\hline
\noalign{\smallskip}
20020815&&-53.0&&-6.4&-39.3&-5.9&-3.3&-1.6&-2.3&&&& \\
20141018&-12.5&-89.9&21.8&-18.6&-17.9&-8.7&-5.7&-2.8&-2.7&-15.8&-9.0&-9.8&-5.6 \\
\hline \hline
\end{tabular}
\endgroup

\begin{multicols}{2}

The long-term light curves and color indexes of V2 gives grounds to predict different reasons for observed variability of the star $-$ existence of hot and cool spots on the stellar surface, and irregular obscuration of the star by circumstellar material. The Fig. 11 shows evidences for color reversal, especially for the $V-R$ and $B-V$ indexes. The color reversal supports the assumption that at least one of the reasons for the observed declines in the star's brightness is obscuration of the star by circumstellar material.

Our spectral observations of V2 revealed an extensive emission spectrum, as only the sodium doublet is observed in absorption (Table 3).
The emission lines of hydrogen, iron, helium and forbidden lines of oxygen dominate the spectrum of the star.
All emission lines show a strong variability of intensity and of full width at half-maximum. 
Therefore, the star has photometric and spectral characteristics of CTTS.

%------------------------------------------------------------------------------------

\subsection{3.4 V3 (2MASS J21403852+6635017, NGC 7129 S V3)}

The star 2MASS J21403852+6635017 (hereafter V3) was reported as a variable star in Semkov (2000). V3 is part from a visual double system and it is located at $\sim$1.2 arc mininutes from V391 Cep.

Kun et al. (2009) measured $I$ = 15.34, $R$ = 16.45, $V$ = 17.78 and $B$ = 19.55 magnitudes for V3 and $I$ = 16.77 mag for the second component. The distance between the two components determined by Kun et al. (2009) is 3.2 arc seconds. T
he authors defined the spectral class of V3 as K5 and determined its mass as 1.15 M$_{\odot}$, its effective temperature as 4350 K, and its age as 2.5 Myr. The spectrum of V3 show strong H$\alpha$ emission line and the star is classified as a CTTS.

   \end{multicols}
   \includegraphics[width=12cm, angle=0]{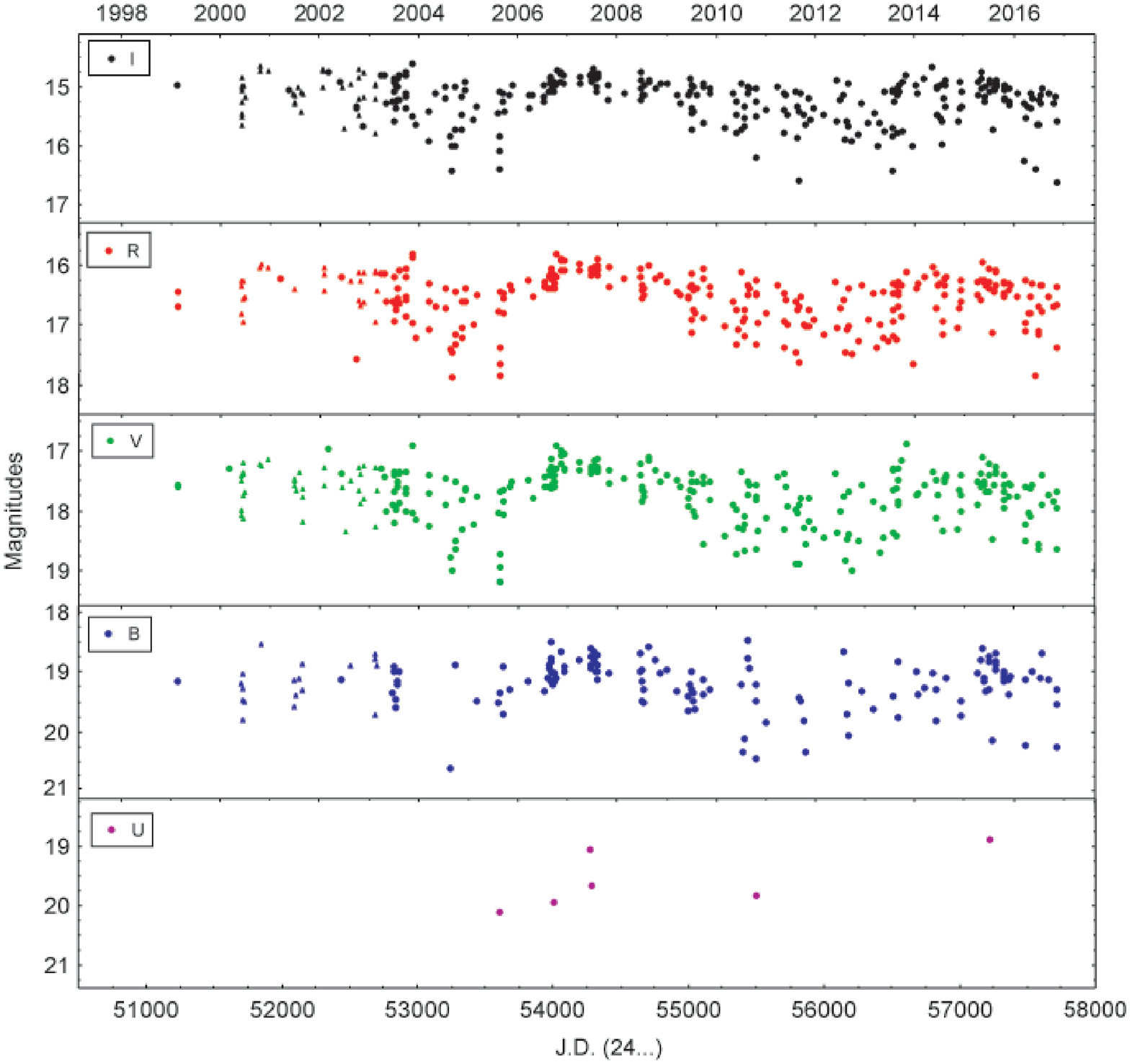}
    \figurecaption{12. }{CCD $UBVRI$ light curves of V3 for the period February 1999$-$November 2016.}
   \begin{multicols}{2}

Fig. 12 present $UBVRI$ light curves of V3 from all our CCD observations (Semkov (2003b) and the present paper), obtained in the period 1999$-$2016. The symbols used are as in Fig. 8. The results of our long-term multicolor CCD observations of the star V3 will be accessible through the CDS database. The average value of the errors in the measured magnitudes are $0.01$-$0.03$ for the $I$- and $R$-band data, $0.01$-$0.03$ for the $V$-band data, $0.02$-$0.06$ for the $B$-band data, and $0.07$-$0.13$ for the $U$-band data. 	

The star shows a large amplitude photometric variability in all bands. It can be seen from Fig. 12 that V3 usually spends most of the time at maximal light. During our photometric monitoring deep declines with different amplitudes in the star's brightness are registered. It is possible to suggest the existence of frequent deep fading events during periods with insufficient data.

The brightness of V3 during the period of our CCD observations 1999$-$2016 varies in the range 14.61$-$16.60 mag for the $I$-band, 15.81$-$17.87 mag for the $R$-band, 16.88$-$19.20 mag for the $V$-band, 18.46$-$20.44 mag for the $B$-band and 18.90$-$20.13 mag for the $U$-band. Because the limit of our photometric data we have only six photometric points in the $U$-band. 
In a very low light the brightness of the star in $B$-band is under the photometric limit of the 60-cm Cassegrain and the 50/70-cm Schmidt telescopes. Evidences of periodicity in the brightness variability of V3 are not detected.

The measured color indexes $V-I$, $V-R$ and $B-V$ versus stellar $V$ magnitude for the period of all our CCD observations are plotted in Fig. 13.
It can be seen from the figure that real color reverse is registered. 
The observed drops in brightness probably are caused by clouds of dust orbiting the star, dense enough to cause UXor phenomenon. 

	 \end{multicols}
   \begingroup\centering
   \includegraphics[width=3.5cm, angle=0]{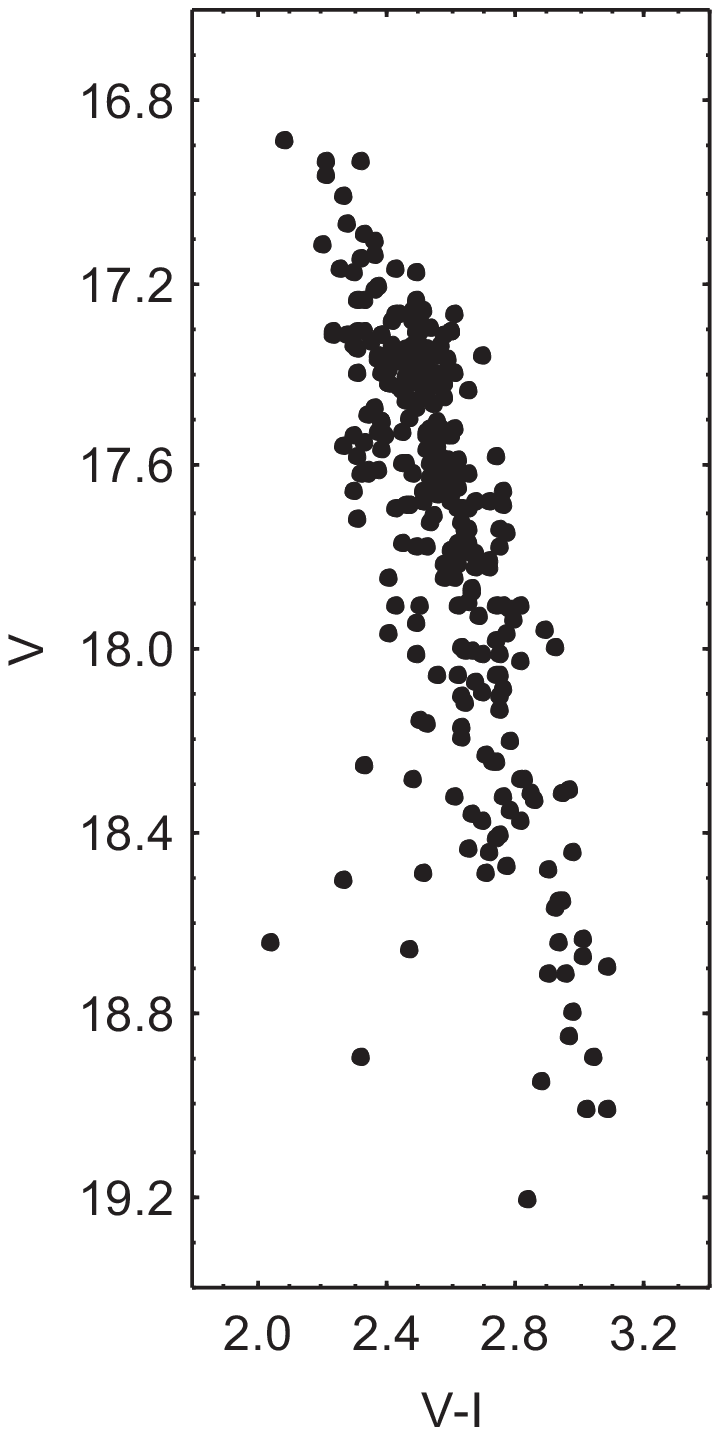}
	 \includegraphics[width=3.5cm, angle=0]{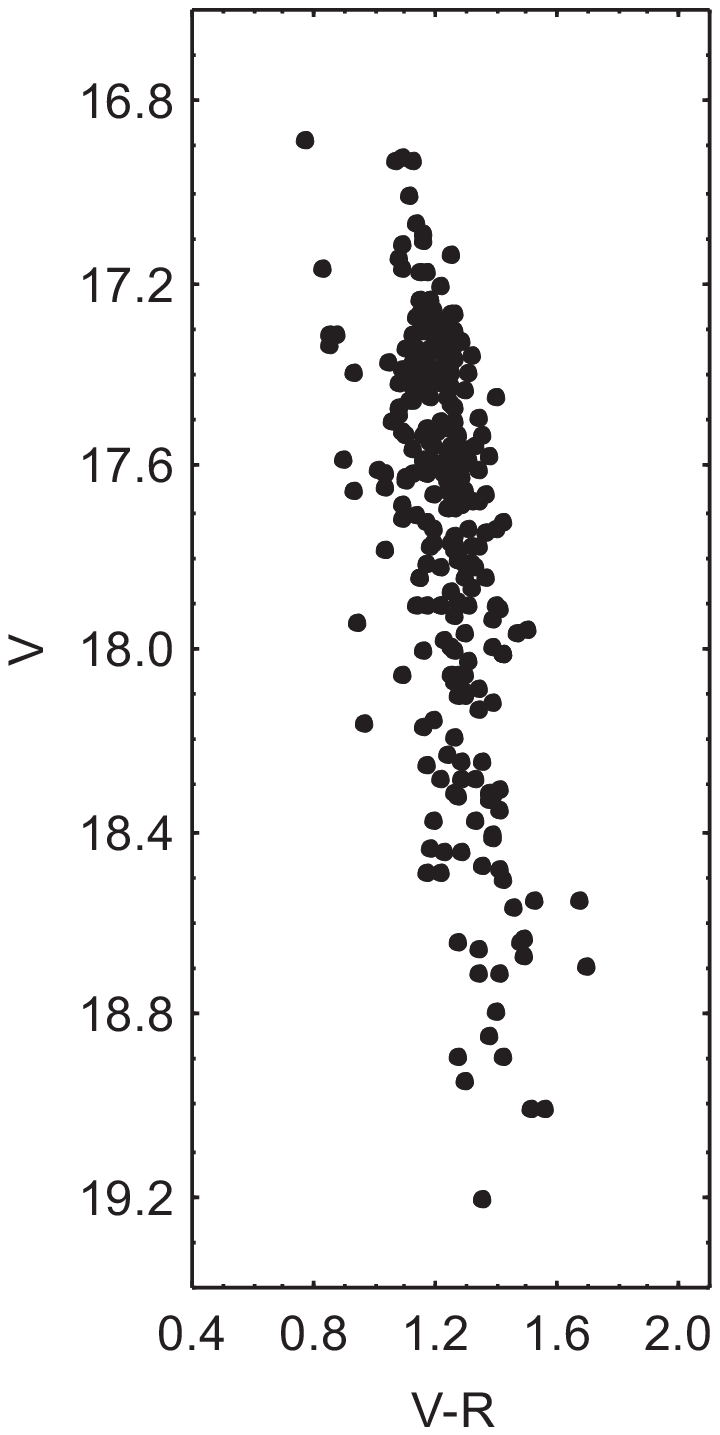}
   \includegraphics[width=3.5cm, angle=0]{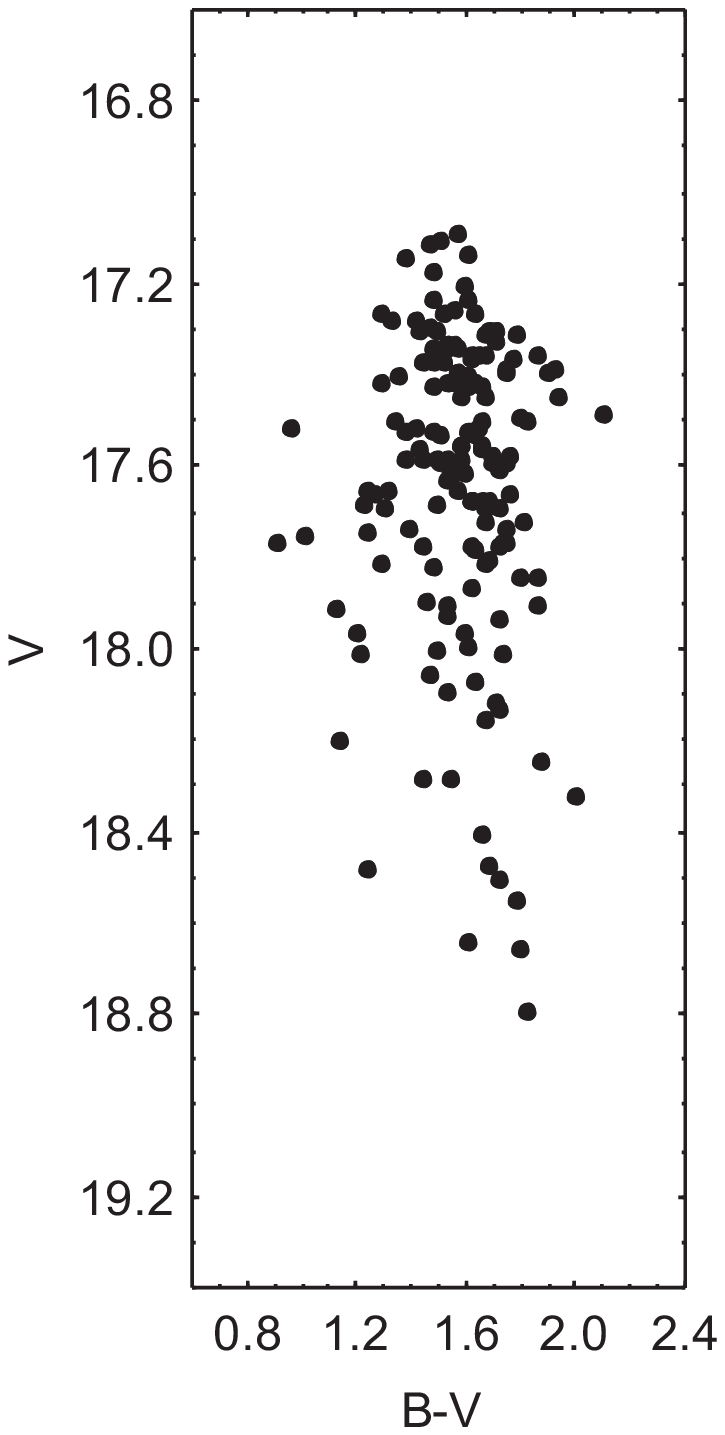}
    \figurecaption{13. }{The $V-I$, $V-R$ and $B-V$ color indexes versus $V$ magnitude for V3 in the period of all our CCD observations.}
 \endgroup

\vskip.5cm \noindent
\parbox{\textwidth}{
{\bf Table 4. }{Spectral lines identified in the optical spectra of V3 and lines' equivalent width.}}  
	\begin{tabular}{llllllllllllll}
\hline \hline
\noalign{\smallskip}
Date&Fe I &[O I] &NaD&[O I] &H$\alpha$&Fe I &Fe II &Fe I&Fe I &He I &Fe II &Si III  \\
yyyymmdd& 5461 &  5577 & 5890 & 6300 &6563 & 6829 & 6862 &6948 & 7248& 7340 &7369 \\ 
\hline
\noalign{\smallskip}
20010907&-1.8&-32.5&-6.0&-11.1&-15.4&-2.9&&-1.2&&&& \\
20141018&-10.6&-64.6&-13.2&-8.3&-27.7&-5.2&-4.1&-2.4&-8.7&-5.6&-6.0&-4.0 \\
\hline \hline
\end{tabular}

	\begin{multicols}{2}

Our spectral observations of V3 revealed an extensive emission spectrum without absorption lines (Table 4).
The emission lines of hydrogen, iron, sodium and forbidden lines of oxygen dominate the spectrum of the star.
All emission lines show a strong variability of intensity and of full width at half-maximum. 
Therefore, the star shows spectral characteristics of CTTS and photometric characteristics of both CTTS and UXor.

%------------------------------------------------------------------------------------

\subsection{3.5 V4 (2MASS J21403576+6635000)}

The star 2MASS J21403576+6635000 (hereafter V4) is discovered as a variable during the present study. 
The star is located at coordinates RA$_{J2000}$ = 21$^{h}$ 40$^{m}$ 35.77$^{s}$ and Dec$_{J2000}$ = +66$^{\circ}$ 35' 01",  $\sim$50 arc seconds from V391 Cep. 
V4 shows very strong and fast photometric variability during short time periods (several minutes or hours) with large amplitude.

The photometric results from our long-term CCD observations of V4 will be accessible through the CDS database. The average value of the errors in the measured magnitudes are $0.01$-$0.02$ for the $I$-band data and $0.01$-$0.10$ mag for the $R$-band data. The $RI$ light curves of V4 from our photometric observations are shown in Fig. 14. 

   \end{multicols}
   \includegraphics[width=13cm, angle=0]{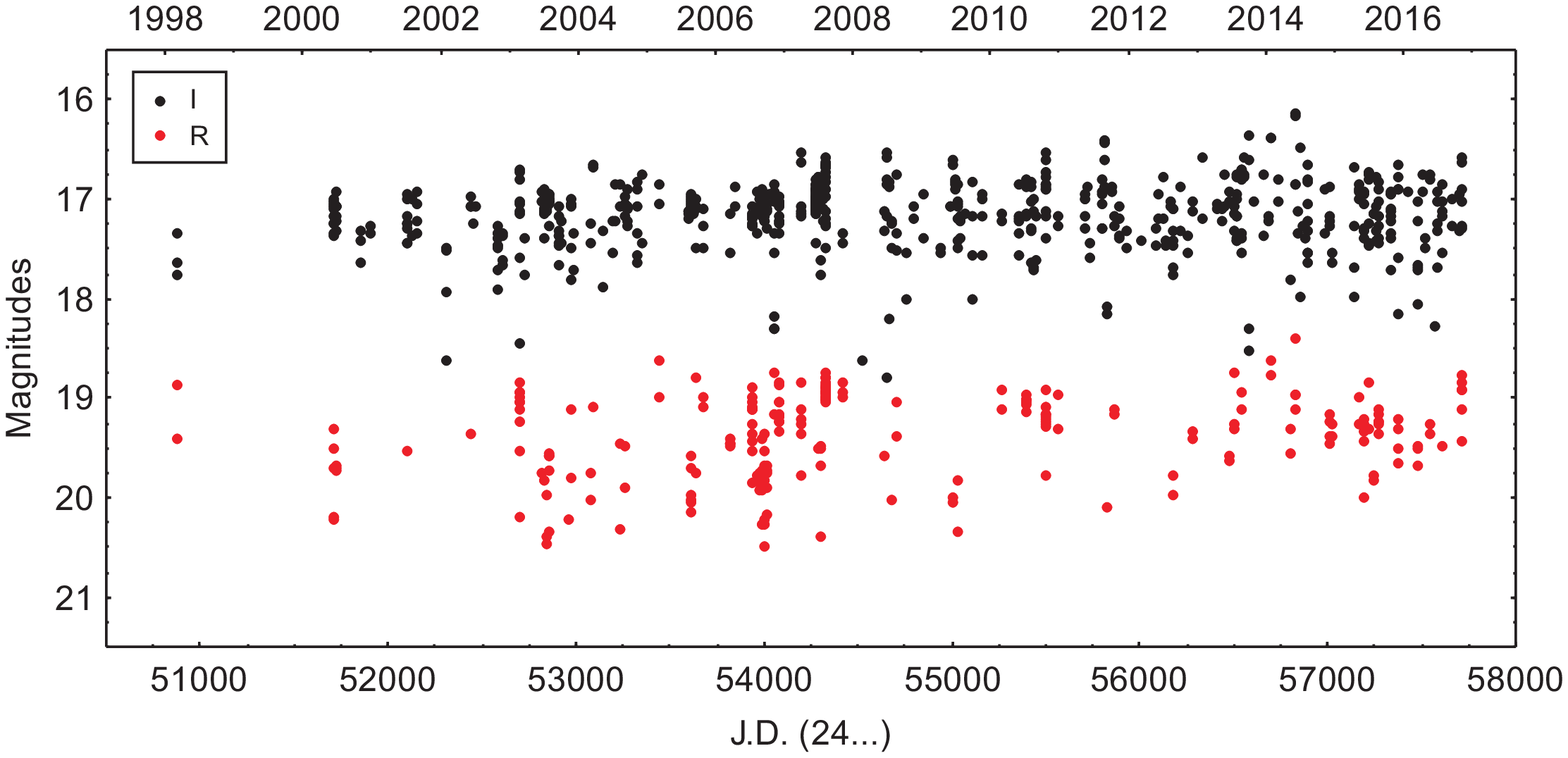}
   \figurecaption{14. }{CCD $UBVRI$ light curves of V4 for the period February 1998$-$November 2016.}
   \begin{multicols}{2}
	
The brightness of V4 during the period of our photometric observations 1998$-$2016 varies in the range 16.14$-$18.79 mag for the $I$-band and 18.40$-$20.48 mag for the $R$-band. The observed amplitudes are $\Delta I$ = 2.65 mag and $\Delta R$ $\sim$ 2.08 mag in the same period. Because of the limit of our photometric data the deeper drops in the brightness of the star in $R$-band and photometric data in $V$- and $B$-bands are not registered.
The photometric monitoring for short periods shows strong variability within a few hours (Fig. 15).
Evidences of periodicity in the brightness variability of V4 are not detected.
The star appeared to be too faint for spectral observations with the 1.3-m RC telescope.

It can be seen from Fig. 14 that during our study the brightness of V4 vary around some intermediate level. 
The presence of V4 in the field of star formation and the irregular variability with large amplitude suggesting for a PMS nature of the star.  
Also, from Fig. 2 can be seen that V4 have strong infrared excess $-$ an indication for the presence of a disk surrounding the star. 
We suggest that the V4 is a TTS and the observed fast variability with large amplitudes probably is caused from strong irregular accretion rate from circumstellar disk onto stellar surface.

   \end{multicols}
   \includegraphics[width=13cm, angle=0]{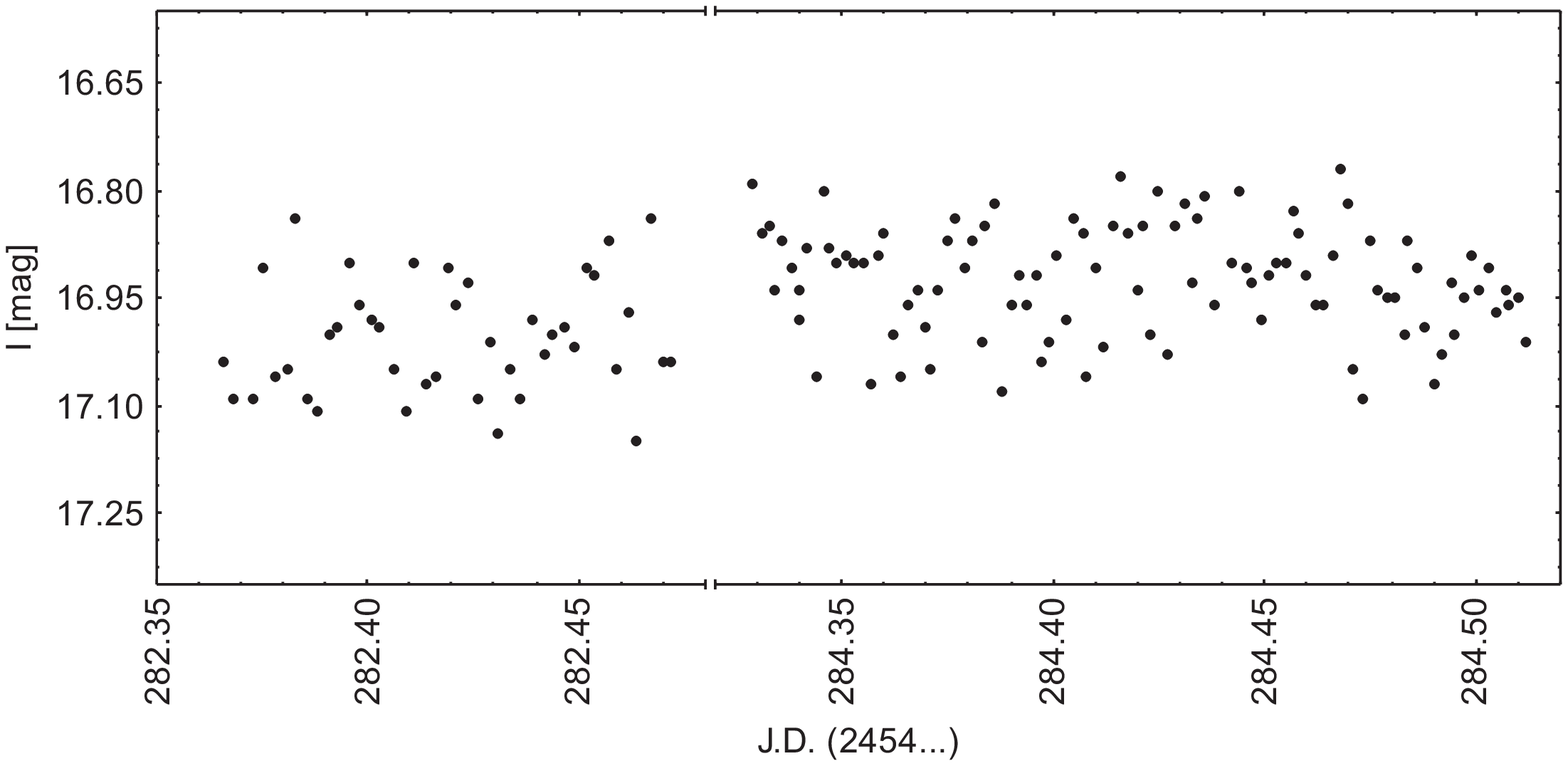}
   \figurecaption{15. }{CCD $I$ light curve of V4 for June 30 and July 02, 2007.}
   \begin{multicols}{2}

\section{4. CONCLUSIONS}

The photometric and spectral results presented in this study show that in the area around V391 Cep is placed processes of intensive star formation. 
All monitored objects showed very strong photometric variability and emission spectra characteristic of CTTS.
V391 Cep, V2 and V3 form a small group of PMS objects with masses close to the solar mass and they probably were formed at about the same time.
The stars are relatively near each other in the vicinity of the dark cloud of interstellar gas and dust.
The another star from our study V1 lies at a distance from the others PMS objects and probably it was formed beyond this group.

Special attention in future studies must be paid to the star V4.
This PMS object belongs to the group around V391 Cep and it shows variability in brightness with extremely large amplitude.
Future spectral observations of V4 could confirm or reject our suggestion about its PMS nature.
Howsoever, our multicolor photometric monitoring of the PMS objects in this area shows the usefulness of long term photometric studies of TTS.

% Acknowledgements

\acknowledgements{The authors thank the Director of Skinakas Observatory Prof. I. Papamastorakis and Prof. I. Papadakis for the award of telescope time. 
This work was partly supported by the Bulgarian Scientific Research Fund of the Ministry of Education and Science under the grants DN 08-1/2016, and DN 18-13/2017 as well as by the project RD-08-37/2019 of the University of Shumen. 
This research has made use of the NASA's Astrophysics Data System Abstract Service, the SIMBAD database and the VizieR catalogue access tool, operated at CDS, Strasbourg, France. This publication makes use of data products from the Two Micron All Sky Survey, which is a joint project of the University of Massachusetts and the Infrared Processing and Analysis Center/California Institute of Technology, funded by the National Aeronautics and Space Administration and the National Science Foundation (Skrutskie et al. 2006).}

% References

\references

{\'A}brah{\'a}m, P., Bal{\'a}zs, L. G. and Kun, M.: 2000, \journal{Astron. Astrophys}, \vol{354}, 645.

Barsunova, O. Yu., Grinin, V. P., Arharov, A. A., Semenov, A. O., Sergeev, S. G. and Efimova, N. V.:	2016, \journal{Astrophysics}, \vol{59}, 147.	

Bessell, M. S.: 1979, \journal{Publ. Astron. Soc. Pac.}, \vol{91}, 589.

Bessell, M. S. and Brett, J. M.: 1988, \journal{Publ. Astron. Soc. Pac.}, \vol{100}, 1134.

Bibo, E. A. and Th\'{e}, P. S.: 1990, \journal{Astron. Astrophys}, \vol{236}, 155.
 
Bouvier, J., Chelli, A., Allain, S., et al.: 1999, \journal{Astron. Astrophys}, \vol{349}, 619.

Bouvier, J., Alencar, S. H. P. and Boutelier, T.: 2007, \journal{Astron. Astrophys}, \vol{463}, 1017.

Cardelli, J. A., Clayton, G. C. and Mathis, J. S.: 1989, \journal{Astrophys. J.}, \vol{345}, 245.

Carpenter, J. M.: 2001, \journal{Astron. J.}, \vol{121}, 2851.

D'Alessio, P., Calvet, N., Hartmann, L., Lizano, S. and Canto, J.: 1999, \journal{Astrophys. J.}, \vol{527}, 893.

Grinin, V. P., Kiselev, N. N., Minikulov, N. Kh., Chernova, G. P. and Voshchinnikov, N. V.: 1991, \journal{Astrophys. Space Sci.}, \vol{186}, 283.

Joy A. H.: 1945, \journal{Astrophys. J.}, \vol{102}, 168.

Hartigan, P. and Lada, C. J.: 1985, \journal{Astrophys. J. Supl.}, \vol{59}, 383.

Herbig, G. H.: 1962, \journal{Adv. Astron. Astrophys.}, \vol{1}, 47.

Herbst, W., Herbst, D. K., Grossman, E. J. and Weinstein, D.: 1994, \journal{Astrophys. J.}, \vol{108}, 1906.

Herbst, W., Eisloffel, J., Mundt, R. and Scholz, A.: 2007, in "Protostars and Planets V", eds. B. Reipurth, D. Jewitt and K. Keil, University of Arizona Press, Tucson, 297.

Ibryamov, S., Semkov, E. and Peneva, S.: 2014, \journal{Res. Astron. Astrophys.}, \vol{14}, 1264.

Ibryamov, S., Semkov, E. and Peneva, S.: 2015, \journal{Publ. Astron. Soc. Aust.}, \vol{32}, e021.

Ibryamov, S., Semkov, E. and Milanov, T., Peneva, S.: 2017, \journal{Res. Astron. Astrophys.}, \vol{17}, 20.

Kun, M., Bal\'{a}zs, L. G. and T\'{o}th, I.: 1987, \journal{Astrophys. Space Sci.}, \vol{134}, 13.

Kun, M.: 1998, \journal{Astrophys. J. Supl.}, \vol{115}, 59.

Kun, M., Kiss, Z. T. and Balog, Z.: 2008,  in "Handbook of Star Forming Regions", ed. B. Reipurth, \vol{VI}, ASP Monograph Publications, San Francisco, CA, 136.

Kun, M., Balog, Z., Kenyon, S. J., Mamajek, E. E. and Gutermuth, R. A.: 2009, \journal{Astrophys. J. Supl.}, \vol{185}, 451.

Magakian, T. Yu. and Movsessian, T. A.: 1997, \journal{Astron. Rep.}, \vol{41}, 483.

Mathieu, R. D., Walter, F. M. and Myers, P. C.: 1989, \journal{Astron. J.}, \vol{98}, 987.

M\'{e}nard, F. and Bertout, C.: 1999, in "The Origin of Stars and Planetary Systems", eds. V. J. Lada and N. D. Kylafis, Kluwer Academic Publishers, Dordrecht, 341.

Meyer, M. R., Calvet, N. and Hillenbrand, L. A.: 1997, \journal{Astron. J.}, \vol{114}, 288.

Miranda, L. F., Eiroa, C. and Gomez de Castro, A. I.: 1993, \journal{Astron. Astrophys}, \vol{271}, 564.

Reipurth, B. and Zinnecker, H.: 1993, \journal{Astron. Astrophys}, \vol{278}, 81.

Reipurth, B., Lindgren, H., Mayor, M., Mermilliod, J.-C. and Cramer, N.: 2002, \journal{Astron. J.}, \vol{124}, 2813.

Rodriguez, J. E., Zhou, G., Cargile, P. A., et al. 2017, \journal{Astrophys. J.}, \vol{836}, 209.

Schneider, P. C., Manara, C. F., Facchini, S. et al., 2018, \journal{Astron. Astrophys}, \vol{614}, A108.

Scholz, A., Irwin, J., Bouvier, J., Sip\H{o}cz, B. M., Hodgkin, S. and Eisl\"{o}ffel, J.: 2011, \journal{Mon. Not. R. Astron. Soc.}, \vol{413}, 2595.

Semkov, E. H. and Tsvetkov, M. K.: 1986, in "Stars Clusters and Associations", ed. G. Sz\'{e}cs\'{e}nyi-Nagy, Publications of the Astronomy Department of the E\"{o}tv\"{o}s University, Budapest, 141.

Semkov, E. H.: 1993a, \journal{Inf. Bull. Var. Stars}, \vol{3825}, 1.

Semkov, E. H.: 1993b, \journal{Inf. Bull. Var. Stars}, \vol{3870}, 1.

Semkov, E. H.: 1993c, \journal{Inf. Bull. Var. Stars}, \vol{3918}, 1.

Semkov, E. H.: 1996, \journal{Inf. Bull. Var. Stars}, \vol{4339}, 1.

Semkov, E. H., Mutafov, A. S., Munari, U. and Rejkuba, M.: 1999, \journal{Astron. Nachr.}, \vol{320}, 57.

Semkov, E.: 2000, \journal{Poster proceedings of IAU Symp. No. 200}, Eds. B. Reipurth and H. Zinnecker, 121.

Semkov, E. H.: 2002, \journal{Inf. Bull. Var. Stars}, \vol{5214}, 1.

Semkov, E. H.: 2003a, \journal{Inf. Bull. Var. Stars}, \vol{5373}, 1.

Semkov, E. H.: 2003b, \journal{Inf. Bull. Var. Stars}, \vol{5406}, 1.

Semkov, E. H.: 2004a, \journal{Inf. Bull. Var. Stars}, \vol{5556}, .1

Semkov, E. H.: 2004b, \journal{Balt. Astron.}, \vol{13}, 538.

Shevchenko, V. S. and Yakubov, S. D.: 1989, \journal{Astronomicheski\v{ı} Zhurnal}, \vol{66}, 718.

Skrutskie, M. F., Cutri, R. M., Stiening, R. et al.: 2006, \journal{Astron. J.}, \vol{131}, 1163.

Sousa, A. P., Alencar, S. H. P., Bouvier, J., et al.: 2016, \journal{Astron. Astrophys}, \vol{586}, A47.

Straiz\v{y}s, V., Maskoli\={u}nas, M., Boyle, R. P. et al.: 2014, \journal{Mon. Not. R. Astron. Soc.}, \vol{438}, 1848.

White, R. J. and Basri, G.: 2003, \journal{Astrophys. J.}, \vol{582}, 1122.

\endreferences

%\end{multicols}

\end{multicols}
\end{document}